# Generative AI and LLMs in Industry: A text- mining Analysis and Critical Evaluation of Guidelines and Policy Statements Across Fourteen Industrial Sectors

**Junfeng Jiao[1] Saleh Afroogh*[2] Kevin Chen[3] David Atkinson[4] Amit Dhurandhar[5]**

1. Urban Information Lab, The University of Texas at Austin, Austin, TX 78712, USA. jjiao@austin.utexas.edu
2. Urban Information Lab, The University of Texas at Austin, Austin, TX 78712, USA Saleh.afroogh@utexas.edu
3. Urban Information Lab, The University of Texas at Austin, Austin, TX 78712, USA. xc4646@utexas.edu
4. Allen Institute for AI (AI2), Seattle, USA davida@allenai.org
5. IBM Research Yorktown Heights, USA, adhuran@us.ibm.com

* Corresponding author: saleh.afroogh@utexas.edu

**Abstract**

The rise of Generative AI (GAI) and Large Language Models (LLMs) has transformed industrial landscapes, offering unprecedented opportunities for efficiency and innovation while raising critical ethical, regulatory, and operational challenges. This study conducts a text-based analysis of 160 guidelines and policy statements across fourteen industrial sectors, utilizing systematic methods and text-mining techniques to evaluate the governance of these technologies. By examining global directives, industry practices, and sector-specific policies, the paper highlights the complexities of balancing innovation with ethical accountability and equitable access. The findings provide actionable insights and recommendations for fostering responsible, transparent, and safe integration of GAI and LLMs in diverse industry contexts.

**Keywords**: Generative AI, Large Language Models, AI governance, industrial applications, ethical AI, policy analysis, text-based analysis, responsible innovation

## Introduction

Generative AI (GAI), particularly Large Language Models (LLMs), is reshaping industries by enabling the creation of content such as text, images, and music through pattern recognition from large datasets. LLMs, as advanced neural network architectures, predict linguistic patterns and generate contextually coherent text, transforming how textual content is created and utilized. These technologies have become integral to industries, driving advancements in automation, customer service, and content generation. In Natural Language Processing (NLP), LLMs represent a significant leap forward. With millions or billions of parameters, they are trained on diverse datasets to replicate complex linguistic structures [1]. Using conditional probability and the chain rule [2], they produce human-like text, characterized by semantic coherence and adaptability. Their applications in industrial processes demonstrate their potential to enhance operational efficiency and innovation [1], [3].

However, their integration comes with ethical, regulatory, and operational challenges. Concerns about data security, misinformation, and algorithmic bias demand comprehensive governance to ensure responsible adoption. While LLMs can enhance workflows, they also risk perpetuating biases and creating accountability gaps if not effectively managed [1]. Addressing these issues requires thoughtful policies tailored to diverse industry contexts.

This study explores the evolving role of GAI and LLMs in industrial sectors, examining their benefits, risks, and governance implications. Through a systematic review of policies and practices, it provides actionable insights and recommendations to align innovation with ethical accountability, fostering responsible and effective deployment across industries.

### 1.2: Escalating significance and prevalence of LLMs and generative Ais in industry settings

Within industry, the integration of LLMs and other GAI technologies has introduced unprecedented opportunities and challenges. These models have shown remarkable capabilities in generating human-like text, aiding in tasks such as customer support, automated content creation, and product recommendations. However, their usage also raises significant operational, ethical, safety and legal considerations that necessitate careful evaluation and guidelines for responsible deployment.[4][5]

The increasing relevance and widespread use of GAI/LLM in industry contexts have prompted diverse responses and adaptations from companies worldwide. According to McKinsey & Company, 21% of companies have developed formal policies governing the use of generative AI in the workplace, and only 32% are addressing the most frequently cited risk: inaccuracy[6] [7]. Many organizations still remain unprepared for the business risks these technologies may bring, such as intellectual property infringement and cybersecurity vulnerabilities.

Moreover, more than 50% of generative AI adopters reportedly use unapproved AI tools in their workplace according to a survey conducted by Salesforce [8] As companies increasingly explore generative AI applications in product development and customer service, a McKinsey survey shows that 33% of organizations are now using AI in at least one business function[6]. Despite the rapid adoption, there are concerns about the lack of comprehensive policies guiding AI use, underscoring the need for industry-level frameworks to address ethical risks and regulatory compliance.

This study seeks to delve into the multifaceted landscape of GAI, with a primary focus on LLMs, within industry settings. Through a careful consideration of the operational implications, ethical considerations, and legal frameworks surrounding the use of these powerful AI tools, this study aims to provide insights and recommendations for navigating the complex intersection of AI technology and industry. Through a systematic evaluation of existing guidelines and practices, we aim to contribute to the development of comprehensive and ethically sound frameworks for the usage of GAIs, including LLMs, in industry contexts.

## II. Methodology

In this research, we undertook a detailed comparative study analyzing data from 160 companies to evaluate workplace policies and guidelines for employing GAIs and LLMs in industrial environments. Our methodology was rigorously structured to ensure an in-depth assessment of these guidelines. We deliberately chose companies that represent a wide range of global institutions, including leading firms from various countries and continents renowned for their industrial prowess. Furthermore, our selection criteria spanned 14 different industrial sectors, such as healthcare, finance, and publishing, to encompass a comprehensive array of viewpoints. Moreover, while the prevailing focus of guidelines pertains to LLMs, this investigation endeavors to explore overarching principles pertinent to all generative AIs, with the objective of fostering inclusivity. The dataset provided for this study is available online on Harvard Dataverse: IGGA: A Dataset of Industrial Guidelines and Policy Statements for Generative Ais.[9],[10],[11].

In exploring the complex landscape of industrial policies governing Large Language Models (LLM) and generative AI, it is crucial to employ effective categorization criteria to systematically analyze diverse aspects. Two primary criteria emerge as particularly pertinent for this research: Geographic Location and Industry Sectors. The choice of Geographic Location allows for a comprehensive understanding of how different countries and regions approach the regulation, promotion, and professional considerations associated with LLM and generative AI. This criterion facilitates the identification of regional variations in policies, considering diverse socio-political contexts and regulatory frameworks. Simultaneously, focusing on Industry Sectors recognizes the varied applications and challenges of these technologies. Different sectors may require tailored policies, given the unique professional concerns and practical implications of implementing LLM and generative AI in specific industries. Together, these criteria provide a structured approach to unravel the nuanced dynamics of industrial policies, offering insights into both geographical diversity and sector-specific considerations.

Moreover, navigating the intricate landscape of industrial policies for Large Language Models (LLM) and generative AI involves a detailed examination of industry sectors. In healthcare, subcategories encompass pharmaceuticals, medical research, and healthcare technology, focusing on tailored policies addressing challenges like ethical and safety considerations and data security. Finance and banking categories include banking, financial technology (fintech), and investment sectors, where policies are analyzed for data security, risk management, and professional considerations in financial applications. In technology and IT services, subcategories cover software development, cloud computing, and data analytics, exploring policies related to innovation, intellectual property, and responsible AI practices. This structured approach enables a comprehensive analysis, offering insights into how different regions and industries shape and respond to LLM and generative AI policies. Sector-specific nuances allow for meaningful cross-sectoral comparisons, revealing diverse regulatory frameworks and their impact on technology deployment and professional considerations. For example, variations may arise in healthcare policies regarding the trustworthy use of generative AI in medical research, and finance policies may differ between traditional banking and fintech applications concerning data security.

In the initial phase of our data collection, we concentrated on identifying top-tier companies across various sectors that had established official guidelines specifically addressing the deployment of GAI and LLMs. This focused approach was chosen to ensure methodological consistency and to maintain the reliability of our analysis. When official guidelines were absent, we included documents found on company websites that articulated their policies on using LLMs, which we treated as policy statements or workplace policies. Additionally, when there was no official guideline or policy available on their websites, interviews with company heads featured in prominent media were also considered as valid sources of information. Companies without any form of guidelines or policy statements were intentionally omitted from our survey to uphold the integrity of our findings. To offset these exclusions, we meticulously selected alternative companies that represented a diverse array of 14 different industrial sectors from various countries and continents, ensuring a balanced and representative sample that enhances the credibility and relevance of our results.

Following the data collection phase, our review process entailed a systematic categorization and in-depth analysis of the guidelines from the chosen companies. Each set of guidelines was thoroughly examined for its scope, specific recommendations, ethical and safety considerations, and implications for using LLMs and GAI in business practices. This detailed scrutiny allowed us to delineate key themes, identify prevailing trends, and pinpoint areas where practices diverged across different geographic and industrial contexts. This comprehensive examination provided us with a nuanced understanding of the current landscape regarding corporate guidelines on GAI and LLM usage.

We conducted a text mining-based analysis of the artificial intelligence usage guidelines collected from 160 company documents spanning 14 distinct industries. This analysis began with tokenizing the collected text data, where each document was first divided into sentences and then further segmented into individual words. This granularity enabled detailed scrutiny of the text. For tokenization, we utilized the Natural Language Toolkit's (nltk) sent_tokenize and word_tokenize functions. To focus on substantive content, we removed stopwords—commonly used words with minimal semantic impact—using the nltk's stopwords list, reducing noise in the dataset and emphasizing meaningful content.

Following tokenization, we applied stemming and lemmatization techniques. Stemming, executed through nltk's PorterStemmer, reduced words to their base form, thus consolidating different inflections of the same word. Lemmatization, a more context-sensitive method using WordNetLemmatizer, transformed words into their dictionary forms, aiding in normalizing the text. Together, these steps enhanced the reliability of word frequency counts.

With the cleaned and normalized dataset, we then constructed a Term Frequency-Inverse Document Frequency (TF-IDF) model using sklearn's TfidfVectorizer. This model evaluated word relevance relative to each document, balancing the term's frequency within the document (Term Frequency) against its distribution across the dataset (Inverse Document Frequency). The TF-IDF model identified distinctive and contextually significant terms within the guidelines, highlighting industry-specific terms. Further, using the KMeans algorithm, we clustered these terms based on their TF-IDF values, revealing prominent themes. Finally, clustering visualizations illustrated the distribution and importance of specific terms across industry sectors.

**Table 1.** Workflow of the analysis of GAI/LLM industrial guidelines and policy statements

| Analytics Workflow | Description | |
|---|---|---|
| Data preparation | Focused on selecting 160 top-tier companies across 14 industrial sectors globally | ensuring a diverse representation of countries and regions. |
| Data preprocessing | Excluded companies without official guidelines or policy statements | supplemented data with interviews or company statements when official policies were absent. |
| Systematic review | Categorized and thoroughly analyzed guidelines to evaluate scope | ethical considerations, and regional/sectoral implications for using GAIs and LLMs |
| Qualitative analysis | Identified major themes, concerns, and divergences in guidelines across sectors | emphasizing region-specific and industry-specific challenges |
| | Analyzing text through tokenization using NLTK's sent_tokenize and word_tokenize functions.Filtering out stopwords to focus on substantive text. | |
| Quantitative Exploration | Constructed a TF-IDF model to evaluate the relative importance of terms across industry guidelines and identify sector-specific keywords. | |
| Text mining-based Analysis | Constructing a TF-IDF model using sklearn's TfidfVectorizer to assess word importance relative to documents. | |
| | Applied KMeans clustering to identify and visualize common patterns and themes across guidelines | Utilizing clustering visualizations to highlight differences |

## III. Qualitative Findings and Resultant Themes

The integration of Large Language Models (LLMs) and Generative Artificial Intelligence (GAI) has precipitated a significant paradigm shift not only within academic realms [12][12], [13] but across various industrial sectors as well. As these technologies carve deeper inroads into domains such as healthcare, finance, education, and legal services, they manifest a dual-edged spectrum of profound opportunities and formidable challenges. This juxtaposition of potential and peril underscores the necessity for a rigorous examination of how different industries govern the deployment of these advanced tools through professional guidelines and policies.

The advent of LLMs and GAIs has been met with varying degrees of enthusiasm and trepidation across sectors. In industries like finance, healthcare, and technology, there is a palpable excitement about the potential of these technologies to revolutionize efficiency, enhance decision-making, and drive innovation. Conversely, in sectors such as manufacturing and utilities, the integration of AI technologies is approached with more caution. Concerns center around issues such as job displacement, data security, and the erosion of essential human oversight. In these contexts, professional guidelines are crafted with a focus on preserving workplace integrity and ensuring that technological advances do not outstrip the regulatory landscape.

In navigating the evolving landscape of integrating Generative AI (GAI) and Large Language Models (LLM) within industry settings, companies play a pivotal role in establishing and implementing guidelines that govern their responsible use. Recognizing the transformative impact of these technologies on operations, decision-making, and innovation, companies worldwide face the dual challenge of maximizing their benefits while managing potential risks. This section examines a diverse range of industry-specific guidelines aimed at regulating GAI/LLM usage across various sectors. Based on an in-depth analysis of 160 official guidelines from companies spanning 14 different industries across seven continents, this study offers a comprehensive view of the global landscape of AI governance in industry (See Table 2).

**Table 2: Industrial guidelines and policy statements for LLMs and generative Ais**

| | **Industry** | **Company** | | **Name of document/website** | **Country** | **Continent** |
|---|---|---|---|---|---|---|
| 1 | Healthcare Counseling (10) | 1 | Becker's Healthcare | Should health systems regulate the use of ChatGPT?[14] | USA | North America |
| | | 2 | Orbita Hospitals and Health Care | OpenAI and ChatGPT: A Primer for Healthcare Leaders[15] | USA | |
| | | 3 | Chugai Pharmaceutical Co | Chugai DX Meeting[16] | Japan | Asia |
| | | 4 | Daiichi Sankyo Company | Daiichi Sankyo's Challenge to Realize: 2030 Version[17] | Japan | |
| | | 5 | Hardian Health | How to get ChatGPT regulatory approved as a medical device[18] | UK | Europe |
| | | 6 | AstraZeneca | AstraZeneca data and AI ethics[19] | UK | |
| | | 7 | Procaps Group | Procaps participate in panels on digitalization and artificial intelligence in pharma manufacturing[20] | Colombia | South America |
| | | 8 | 1DOC3 | 1DOC3: ACCESSIBLE HEALTHCARE TO MILLIONS[21] | Colombia | |
| | | 9 | CSL Limited | Artificial Intelligence at CSL[22] | Australia | Australia |
| | | 10 | DokiLink | Artificial Intelligence in Africa's Healthcare: Ethical Considerations[23] | South Africa | Africa |
| 2 | Technology and IT Services (30) | 11 | Microsoft | Artificial Intelligence (AI) usage policy for Microsoft[24] | USA | North America |
| | | 12 | Apple | Apple Restricts Use of ChatGPT[25] | USA | |
| | | 13 | Alphabet | Google AI principles[26] | USA | |
| | | 14 | Meta Platforms, Inc. | Meta Guideline to Responsible AI[27] | USA | |
| | | 15 | Amazon | AWS Responsible AI Policy[28] | USA | |
| | | 16 | Oracle | Oracle Generative AI strategy[29] | USA | |
| | | 17 | Adobe | Adobe Generative AI Guideliens[30] | USA | |
| | | 18 | Qualcomm | Qualcomm AI strategy[31] | USA | |
| | | 19 | Salesforce | Generative AI: 5 Guidelines for Responsible Development[32] | USA | |
| | | 20 | CGI Inc | CGI optimizing generative AI potential[33] | Canada | |
| | | 21 | LG Electronics | LG Presents 'AI Ethics Principles' for Trustworthy AI Research[34] | South Korea | |
| | | 22 | Fujitsu | Fujitsu Generative AI use guidelines[35] | Japan | |
| | | 23 | Panasonic | Panasonic Responsible AI use[36] | Japan | |
| | | 24 | Sony | Sony Responsible AI Usage[37] | Japan | |
| | | 25 | Asm Pacific Technology | ASMPT 2023 Interim Report[38] | Hong Kong | Asia |
| | | 26 | Infosys | Infosys Responsible AI[39] | India | |
| | | 27 | Samsung Group | Samsung Group Digital Responsibility[40] | South Korea | |
| | | 28 | Sea Limited (Garena) | Sea Limited – Digital Transformation Strategies[41] | Singapore | |
| | | 29 | Grab | Grab's AI Ethics Principles[42] | Singapore | |
| | | 30 | Sea Group | Sea Founder Warns of Turmoil From the Shift to AI[43] | Singapore | |
| | | 31 | Atos | Atos blueprint for generative AI[44] | France | Europe |
| | | 32 | Capgemini | AI code of ethics.[45] | France | |
| | | 33 | Dormakaba | AI based anti-tailgating Solution[46] | Switzerland | |
| | | 34 | Ericsson | trustworthy AI[47] | Sweden | |
| | | 35 | Accenture | AI ethics & governance[48] | Ireland | |
| | | 36 | SAP | SAP AI ethics[49] | Germany | |
| | | 37 | amadeus it | Capturing the power of Generative Artificial Intelligence to enhance the passenger experience[50] | Spain | South America |
| | | 38 | Gorilla Logic | Ai in product development[51] | Costa Rica | |
| | | 39 | Atlassian | Atlassian Intelligence is built on trust[52] | Australia | Australia |
| | | 40 | Flutterwave | Flutterwave strategic agreement[53] | Nigeria | Africa |
| 3 | Finance and Banking (10) | 41 | JPMorgan Chase | JPMorgan Chase Restricts Staffers' Use Of ChatGPT[54] | USA | North America |
| | | 42 | Wells Fargo | Wells Fargo, artificial intelligence, and you[55] | USA | |
| | | 43 | Mizuho Financial Group | Mizuho permits 45,000 employees to use generative AI[56] | Japan | Asia |
| | | 44 | State Bank of India | SBI Embraces AI and ML Technologies to Transform Banking Operations[57] | Japan | |
| | | 45 | HSBC Holdings Plc | HSBC principles for ethical use of data and AI[58] | UK | Europe |
| | | 46 | Lloyds Banking Group | How AI automation is helping[59] | UK | |
| | | 47 | Itaú Unibanco Holding | Brazil's Itaú in 'very good position' to harness generative AI[60] | Brazil | South America |
| | | 48 | Banco de la Nacion Argentina | Ethical and responsable use of AI in Argentina worker's rights[61] | Argentina | |
| | | 49 | Westpac | How AI will shape the future of banking[62] | Australia | Australia |
| | | 50 | Standard Bank Group | The Future of Digital Banking and Transacting in Africa[63] | South Africa | Africa |
| 4 | Publication Industry (10) | 51 | Elsevier | The use of generative AI and AI-assisted technologies in writing for Elsevier[64] | USA | North America |
| | | 52 | The New York Times Company | AI workplace changes[65] | USA | |
| | | 53 | China Daily | Guidelines establish proper uses of AI in research[66] | China | Asia |
| | | 54 | South China Morning Post | China unveils new artificial intelligence guidelines [67] | China | |
| | | 55 | Penguin Random House | Penguin Random House CEO hopes AI will help sell more books: Report[68] | UK | Europe |
| | | 56 | Wolters Kluwer | Artificial Intelligence (AI) Principles[69] | Netherland | |
| | | 57 | Grupo Planeta | AI policy[70] | Spain | South America |
| | | 58 | Grupo Editorial Record | Principles for AI[71] | Brazil | |
| | | 59 | Allen & Unwin | AI and International security[72] | Australia | Australia |
| | | 60 | News24 | News24 to use AI in moderating comments[73] | South Africa | Africa |
| 5 | Language Translation Services (10) | 61 | American Translators Association | ChatGPT for Translators: How to Use the Tool to Work More Efficiently?[74] | USA | North America |
| | | 62 | TransPerfect | TransPerfect generative AI[75] | USA | |
| | | 63 | SEAtongue | Using AI for interpretation[76] | Malaysia | Asia |
| | | 64 | Jinyu Translation | Translation: Measures for the Management of Generative Artificial Intelligence[77] | China | |
| | | 65 | SDL plc | SDL to expand knowledge discovery and intelligent process automation to additional languages[78] | UK | Europe |
| | | 66 | Lionbridge | Lionbridge usage of AI[79] | Ireland | |
| | | 67 | Altura Interactive | ChatGPT translation[80] | Mexico | South America |
| | | 68 | GLOBO Brazil | Globo Pacts with Google Cloud in Bid to Become a Mediatech Company[81] | Brazil | |
| | | 69 | Straker Translations | Traker translations security[82] | New Zealand | Australia |
| | | 70 | Elite Translations Africa | Harness the power to AI for preservation of African languages[83] | Kenya | Africa |
| 6 | Construction and Urban Planning (10) | 71 | Bechtel | Applications of Artificial Intelligence In EPC[84] | USA | North America |
| | | 72 | Turner Construction Company | AI at 2023 summit[85] | USA | |
| | | 73 | Hyundai E&C | Hyundai E&C ensures Safety and Quality Management of construction sites[86] | South Korea | Asia |
| | | 74 | Shimizu Corporation | Shimizu develops AI systems for initial structural designs[87] | Japan | |
| | | 75 | Vinci SA | Innovation and prospective[88] | France | Europe |
| | | 76 | Bouygues Construction | Innovation strategy[89] | France | |
| | | 77 | Grupo ACS | Shareholder's meeting & future prospects[90] | Spain | South America |
| | | 78 | Sacyr | IntegratedSustainabilityReport 2023 – Committed to a Sustainable Future[91] | Argentina | |
| | | 79 | Lendlease | Lendlease CEO on using AI in their businesses[92] | Australia | Australia |
| | | 80 | Arab Contractors | UAE National strategy for generative AI[93] | Egypt | Africa |
| 7 | Consulting and management (10) | 81 | McKinsey | About half of McKinsey staff allowed to use generative AI: report[94] | USA | North America |

| | | 82 | Deloitte | Deloitte Launches Innovative ‘DARTbot’ Internal Chatbot[95] | USA | |
|---|---|---|---|---|---|---|
| | | 83 | Tata Consultancy Services | The future is AI. The future is human[96] | India | Asia |
| | | 84 | Infosys | Responsible ai[97] | India | |
| | | 85 | Roland Berger | ChatGPT a game changer for artificial intelligence[98] | Germany | Europe |
| | | 86 | Grupo Assa | Digital transformation practice[99] | Brazil | South America |
| | | 87 | Falconi Consultores de Resultado | Optimizing performance with artificial intlleigence[100] | Brazil | |
| | | 88 | Nous Group | AI powering innovation and productivity [101] | Australia | Australia |
| | | 89 | Africa International Advisors | Future of Africa International Advisors Group[102] | South Africa | Africa |
| 8 | Design and Fashion Technology (10) | 90 | Nike | Nike leveraging AI in operations[103] | USA | North America |
| | | 91 | Ralph Lauren Corporation | Ralph Lauren testing AI[104] | USA | |
| | | 92 | Comme des Garçons | AI designer creating fashion grails from iconic runways[105] | Japan | Asia |
| | | 93 | Shiseido Company, Limited | AI and change in management[106] | Japan | |
| | | 94 | LVMH Moët Hennessy Louis Vuitton SE | Joins stanford developement program[107] | France | Europe |
| | | 95 | Kering | AI and innovation[108] | France | |
| | | 96 | Havaianas | AI powered Case study[109] | Brazil | South America |
| | | 97 | Fatabella | Falabella hires Amelia as a digital assistant for its employees[110] | Chile | |
| | | 98 | Cotton On Group | How Cotton On Is Taking the Aussie Aesthetic Global with AI[111] | Australia | Australia |

| | | 99 | David Tlale | South Africa to adopt AI[112] | South Africa | Africa |
|---|---|---|---|---|---|---|
| 9 | Entertainment and game development (10) | 100 | Activision Blizzard | AI in game development[113] | USA | North America |
| | | 101 | Electronic Arts | AI in game industry[114] | USA | |
| | | 102 | Tencent | Tencent AI policy[115] | China | Asia |
| | | 103 | Square Enix | Letter from the president on AI[116] | Japan | |
| | | 104 | Ubisoft | Ghostwriter using AI in script writing[117] | France | Europe |
| | | 105 | Supercell | Supercell AI fund venture[118] | Finland | |
| | | 106 | Wildlife Studios | AI trust and safety[119] | Brazil | South America |
| | | 107 | Globant | Globant AI Manifesto[120] | Argentina | |
| | | 108 | Wargaming Sydney | AI in Wargaming[121] | Australia | Australia |
| | | 109 | Kukua | Future production with AI[122] | Kenya | Africa |
| 10 | Journalism and News Media (10) | 110 | New York Times | The New York Times is building a team to explore AI in the newsroom[123] | USA | North America |
| | | 111 | ABC News | ABC builds its own AI model[124] | USA | |
| | | 112 | JoongAng Daily | JoongAng Group Builds South Korea's First AI-Driven Enterprise Network by Juniper[125] | South Korea | Asia |
| | | 113 | Asahi Shimbun Company | Panel Discussion Artificial Intelligence and Democracy[126] | Japan | |
| | | 114 | BBC | BBC AI Principles[127] | UK | Europe |
| | | 115 | Reuters | Reuter AI ethics and principles[128] | UK | |
| | | 116 | Grupo Globo | Grupo Globo CEO on Evolving rules and regulations surrounding AI[129] | Brazil | South America |
| | | 117 | Pharu | Pharu and his challenge of bringing analytics to Latin American culture[130] | Chile | |
| | | 118 | News Corp Australia | News Corp AI powered News[131] | Australia | Australia |
| | | 119 | Nation Media Group | State in battle to protect data privacy, enhance security in the AI age[132] | Kenya | Africa |
| 11 | Pharmaceutical Research and Development (10) | 120 | Pfizer Inc. | Pfizer AI policy and position[133] | USA | North America |
| | | 121 | Johnson & Johnson | Jnj policy and positions[134] | USA | |
| | | 122 | Takeda Pharmaceutical Company Limited | Takeda position on AI[135] | Japan | Asia |
| | | 123 | Eisai Co., Ltd. | AI in drug design[136] | Japan | |
| | | 124 | Roche Holding AG | Harnessing the power of AI[137] | Switzerland | Europe |
| | | 125 | Novartis International AG | Ethical and responsible use of AI[138] | Switzerland | |
| | | 126 | EMS S/A | AI in EMS the future is here[139] | Brazil | South America |
| | | 127 | Sanofi | Sanofi Responsible AI principles[140] | Brazil | |
| | | 128 | CSL Limited | Artificial intelligence at CSL[141] | Australia | Australia |
| | | 129 | Fidson | Integrating tech into healthcare profitable – Firm[142] | Nigeria | Africa |
| 12 | Social Media and Networking/ telecommunications (10) | 130 | Twitter (X) | Synthetic and manipulated media policy[143] | USA | North America |
| | | 131 | LinkedIn Corporation | Linkedin engineering responsible AI[144] | USA | |
| | | 132 | Alibaba Group Holding Limited | Alibaba Cloud Unveils New AI Model to Support Enterprises' Intelligence Transformation[145] | China | Asia |
| | | 133 | WeChat | Wechat AI privacy policy[146] | China | Asia |
| | | 134 | Spotify Technology S.A. | Spotify using AI[147] | Sweden | Europe |
| | | 135 | Skype | Skype Translator AI policy[148] | Luxemburg | |
| | | 136 | MercadoLibre, Inc | Improving user experience and boost sales with AI[149] | Argentina | South America |
| | | 137 | América Móvil | The ITU recognizes the Carlos Slim Foundation and América Móvil for their technological innovation in health care[150] | Mexico | |
| | | 138 | Accel IT | Accel's AI Investments Keep The Focus On Applications And Tooling[151] | Australia | Australia |
| | | 139 | MTN SA | MTN SA introduces Siya as it intensifies its AI Strategy in a bid to boost efficiency and revenue[152] | Nigeria | Africa |
| 13 | Advertising and marketing (10) | 140 | WPP plc | AI regulation is about finding the right balance[153] | USA | North America |
| | | 141 | Omnicom Group Inc. | Omnicom first mover access to AI insights[154] | USA | |
| | | 142 | Dentsu Group Inc. | Dentsu AI innovations[155] | Japan | Asia |
| | | 143 | Hakuhodo DY Holdings Inc. | Interview with the CFO on artificial intelligence[156] | Japan | |
| | | 144 | Publicis Groupe SA | PUBLICIS IS PUTTING AI AT ITS CORE TO BECOME THE INDUSTRY'S FIRST INTELLIGENT SYSTEM[157] | France | Europe |
| | | 145 | Havas Group | AI makes its mark at Havas group[158] | France | |
| | | 146 | Grupo ABC | AI resource guide[159] | Brazil | South America |
| | | 147 | DPZ&T | Domino's (DPZ) Boosts AI Capabilities With Microsoft Partnership[160] | Brazil | |
| | | 148 | Clemenger Group | Bad News? Send an AI. Good News? Send a Human[161] | Australia | Australia |
| | | 149 | Ogilvy Africa | Creativity, business and society in the age of AI[162] | South Africa | Africa |

| 14 | Legal Tech/ Legal Services/ Intellectual Property Law (10) | 150 | LegalZoom | LegalZoom Launches Doc Assist in Beta, Combining the Power of GenAI and Our Independent Attorney Network[163] | USA | North America |
|---|---|---|---|---|---|---|
| | | 151 | Thomson Reuters | AI principles[164] | USA | |
| | | 152 | Zegal | Comprehensive impact of AI[165] | Hong Kong | Asia |
| | | 153 | Cyril Amarchand Mangaldas | Legal Technology & Alternative Legal Services Guideline [166] | India | |
| | | 154 | Rocket Lawyer | AI workplace use policy[167] | UK | Europe |
| | | 155 | Allen & Overy | Artificial Intelligence Use Case at A&O[168] | UK | |
| | | 156 | Demarest Advogados | Good practices using AI[169] | Brazil | South America |
| | | 157 | Lopes Pinto, Nagasse | AI, Data Protection & Privacy 2024 legislation[170] | Argentina | |
| | | 158 | Lawpath | Artificial Intelligence: What is it and Can it Help My Business?[171] | Australia | Australia |
| | | 159 | Webber Wentzel | Webber Wentzel embraces Generative AI as part of its ongoing innovation journey[172] | South Africa | Africa |
| | | 160 | ENS Africa | Responsible AI: embracing generative artificial intelligence technologies- a brief guide for organisations[173] | South Africa | |

The results of this exploration are categorized into the following nine subsections, each addressing key aspects of AI regulation and ethical and safe practice within industries (see Table 3). By delving into the strategies employed by industry sectors to navigate the opportunities and challenges posed by GAI/LLM, this section aims to shed light on the multifaceted nature of AI governance within industry and offer insights for fostering responsible innovation and ethical and safe practice.

## 3.1. Health Counseling

Artificial Intelligence (AI) in the healthcare counseling sector has been used to enhance efficiency, better personalizing care, and better maintain patient care. AI technologies, including generative models like ChatGPT, are newly being explored for applications in the healthcare counseling sector, services such as real-time medical advice and streamlining tasks in diagnostics and administrative processes. For example, platforms like 1Doc3 [174] provide personalized, AI-driven healthcare guidance to millions of Spanish-speaking users, better helping patients make better-informed decisions about their healthcare. Furthermore, companies like Chugai Pharmaceutical are utilizing AI to accelerate drug discovery and improve patient outcomes through the development of digital biomarkers [175].

Industry examples further illustrate the application of AI in healthcare. CSL Behring [176] , for example, employs AI to improve patient safety and pharmacovigilance. using natural language processing—a machine learning algorithm—to analyze real-world data and better identify safety signals, thus supporting more accurate and timely healthcare decisions. On the other hand, Procaps Group [177] showcases the use of AI in pharmaceutical manufacturing, where digitalization brings initiatives to improve efficiencies and quality control in medicine, better maintaining high standards in drug production and patient safety.

Despite its potential, integrating AI into healthcare counseling presents challenges, including technical and regulatory hurdles. Hardian Health [178] highlighted regulatory bodies like the FDA and MHRA are developing frameworks to accommodate the continuous updates necessary for AI systems while ensuring compliance. While Dokilink [179] highlights that importance of ethical considerations remain crucial as AI evolves, requiring organizations to adopt ethical guidelines to guide AI development and deployment. This ensures that AI is used responsibly, , focusing on transparency, fairness, and maintaining a human-centric approach.

Currently, AI models like ChatGPT are not yet suitable for medical use. The development of these systems requires a clearer definition of their intended use, as risking the well-being of patients and uncertainties in trusting medical devices have catastrophic consequences. As we see companies bettering their medical services, usage highlights the potential of AI to improve healthcare outcomes and can reduce disparities globally.

## 3.2. Information Technology

The adoption of AI across industries such as telecom, travel, and software development is redefining how big tech and their businesses operate. In the telecom industry, the integration of AI promises enhanced operational efficiency but introduces new risks. In general, large-scale telecom companies emphasize the necessity of trustworthy AI, requiring human agency and oversight so that if AI-controlled systems threaten safety, we can intervene. Transparency through Explainable AI (XAI) techniques make decisions made by AI clearer to understand. Privacy and data ownership with measures like differential privacy being implemented to protect personal data, in line with regulations such as GDPR. Having fallback mechanisms and layered protection against adversarial attacks are being upheld to the highest level by any company training and maintaining reliable and safe AI systems. Many guidelines highlighted establishing AI governance to oversee the ethical design, deployment, and monitoring of AI systems. Regular assessment of AI risks, alongside systematic testing for fairness, transparency, and safety. Managing the broader impacts of AI on the workforce, sustainability, and privacy is also a key concern, aiming to mitigate any potential negative effects while maximizing the benefits AI can offer.

In the software industry, companies like SAP [180] emphasize the ethical use of AI by adhering to established principles, such as those outlined by UNESCO, which focus on transparency, privacy, human oversight, and fairness. These principles guide the development and deployment of AI solutions, ensuring they are aligned with societal values and regulatory requirements. Microsoft's[181] commitment to ethical AI is operationalized through a dedicated steering committee that oversees AI processes, ensuring alignment with ethical standards and continuous engagement with the evolving landscape of AI ethics.

Companies like Amadeus [182] is exploring the transformative potential of Generative AI (GAI) to enhance the traveler experience at every stage, from planning to post-trip interactions. While GAI offers exciting opportunities for personalized and scalable customer engagement, there is a strong focus on implementing guardrails to address potential risks, such as misinformation, bias, and data breaches. Ensuring compliance with regulations, such as the EU AI Act, and adhering to ethical AI principles are crucial for maintaining trust and fostering innovation.

Atlassian's [183] approach to AI reflects a commitment to security, transparency, and scalability. Its AI-powered features are built on a trusted platform that ensures data privacy and respects user permissions. By incorporating responsible technology principles, Atlassian ensures that data is used only for intended purposes, without being shared or used for training models across different customers. Compliance with data protection regulations, such as GDPR, is maintained, and users are provided with controls to manage data usage, ensuring a secure and trusted AI environment.

### 3.3. Finance and Banking

The finance and banking industry is rapidly evolving, largely driven by integrating of artificial intelligence (AI) and machine learning (ML) technologies. Major institutions such as JPMorgan Chase, Wells Fargo, Mizuho Financial Group, State Bank of India (SBI), HSBC, and Itaú Unibanco are increasingly leveraging these technologies to enhance efficiency, improve customer experiences, and ensure compliance with regulatory frameworks.

JPMorgan Chase [184] has restricted the use of ChatGPT among its staff as a precautionary measure to safeguard sensitive financial information, highlighting a growing concern over the security and regulatory implications of AI. Similarly, banks in more underdeveloped countries such as Banco de la Nacion Argentina [185], while embracing the future of AI, are taking precautions to limit the use of OpenAI's tools to protect proprietary data. These moves reflect a broader trend in the financial sector towards cautious adoption of AI, ensuring that security and compliance remain paramount.

Similarly, Itaú Unibanco [186] is also cautiously approaching the adoption of generative AI, focusing on research and development while monitoring regulatory developments. The bank's methodical approach underscores the importance of understanding the implications and risks of AI before full-scale deployment, ensuring that ethical, security, and bias concerns are adequately addressed.

In contrast, Wells Fargo [187] is actively embracing AI to transform its operations, leveraging AI for customer interactions, risk management, and to enhance operational efficiencies. The bank's virtual assistant, Fargo™, powered by Google's conversational AI, is a prime example of how AI can streamline customer service. Wells Fargo's strategic focus on AI is part of a decade-long investment in technology, aiming to integrate AI into various business applications, ensuring alignment with regulatory oversight and maintaining a commitment to responsible technology use.

Mizuho Financial Group [188] is another institution exploring the potential of generative AI. The company is providing access to Microsoft's Azure OpenAI service to thousands of its employees, reflecting a more open approach to AI adoption compared to its peers. Mizuho aims to use AI to improve efficiency and explore innovative solutions, although it remains vigilant about the associated risks, such as privacy concerns and data security.

SBI [189] is leveraging AI and ML to overhaul its banking operations, focusing on areas such as cybersecurity, fraud detection, customer service through chatbots, and credit assessment. The bank's approach illustrates how AI can enhance decision-making processes and operational efficiency, ensuring a safer and more customer-centric banking experience.

HSBC [190] emphasizes the ethical use of AI, guided by its principles to ensure integrity, protect privacy, and maintain transparency. The bank's focus on preventing unfair bias, ensuring accountability, and adapting governance frameworks to meet emerging needs reflects a commitment to using AI responsibly and ethically. This approach aligns with broader industry efforts to develop best practices in the ethical use of AI.

In the broader context, the financial industry is increasingly relying on intelligent automation to improve efficiency and customer service. Institutions like Lloyds Banking Group use robotics and AI to handle repetitive tasks, respond to customer queries, and process large volumes of transactions, demonstrating how automation can support both operational and customer-facing roles. As the use of AI and automation grows, it will be crucial for financial institutions to balance innovation with security, ethical considerations, and regulatory compliance.

### 3.4. Publication Industry

The publishing industry is navigating the rapidly changing landscape brought about by the integration of artificial intelligence (AI), with a strong emphasis on ethical use, transparency, and protection of intellectual property. Major publishing companies such as Elsevier, Penguin Random House, and Harper Collins are exploring AI to enhance their processes and offerings while maintaining stringent guidelines to protect authorship and content integrity.

Elsevier [191], a leading publisher in academic and scientific content, has implemented a policy that guides the use of generative AI and AI-assisted technologies specifically in the writing process. This policy stipulates that while AI can be used to improve the readability and language of works, it should not replace core tasks such as producing scientific insights or drawing conclusions. Authors are required to disclose their use of AI, and Elsevier also prohibits using AI tools for creating or altering images in submitted manuscripts, except when explicitly part of the research design, such as in AI-assisted imaging in biomedical research. This careful approach reflects the broader commitment to uphold ethical standards and accuracy in scholarly publishing.

Penguin Random House [192], another major player in the publishing world, sees AI as a tool to potentially boost book sales and streamline operations. CEO Nihar Malaviya has expressed hope that AI will facilitate the sale of more book titles without necessitating a significant increase in staff. This reflects a strategic focus on growth and efficiency, balancing the integration of new technologies with the need to preserve jobs and maintain high editorial standards. The company's exploration of AI comes in the context of broader industry challenges, including cost-cutting measures and attempts to expand market share, as seen in its bid to acquire Simon & Schuster.

Wolters Kluwer [193]is also looking into AI applications, particularly in producing AI-narrated audiobooks, which could expedite the release of translated novels. Similarly, AI's role in moderating online content is also gaining traction, as seen with News24's [194] plan to reintroduce comments on its platform using AI tools to filter out hateful and discriminatory comments. Such innovations demonstrate how AI is being used enhance accessibility and reach new audiences, albeit with careful consideration to avoid undermining the role of human narrators and translators.

The use of AI in the publishing industry extends beyond text generation to include ethical considerations in content creation and management. The Global Principles for AI, established by Grupo Editorial Record [195] , is a set of guidelines adopted by publishers that emphasizes the importance of intellectual property rights, transparency, and fairness. These principles ensure that AI systems are developed and deployed under frameworks that respect publishers' investments in original content. The principles advocate for the responsible use of AI, ensuring that content is used lawfully and creators are compensated fairly. Transparency in AI processes, from content training to deployment, is crucial for maintaining trust and integrity in published works.

Overall, the publishing industry is harnessing AI to innovate and improve operational efficiency, content accessibility, and user interaction. However, this adoption is tempered with a strong commitment to ethical standards, transparency, and respect for intellectual property, ensuring that AI serves as a tool to augment rather than replace human creativity and judgment.

### 3.5. Language Translation Services

The language translation services industry is evolving rapidly with the integration of artificial intelligence (AI), emphasizing efficiency, accuracy, and ethical considerations. Leading companies such as TransPerfect, SDL, and Lionbridge are leveraging AI to enhance translation processes while maintaining rigorous standards to ensure quality and reliability.

TransPerfect [196], a major player in translation and localization services, has incorporated generative AI into its operations to optimize business processes and reduce costs. The company focuses on three main areas: technology, services, and solutions. TransPerfect's AI.NOW division prioritizes data security and compliance, ensuring that AI-driven translations adhere to strict privacy and confidentiality standards. The company also emphasizes the importance of human oversight in AI applications, using AI to assist rather than replace translators. This approach allows TransPerfect to offer customized solutions that meet the specific needs of its clients, ensuring both efficiency and quality.

SDL [197], known for its AI-based Natural Language Understanding platform, has partnered with Expert System to integrate machine translation capabilities into its offerings. This collaboration aims to create a comprehensive multilingual content understanding platform that supports industries like life sciences and government. SDL's focus on secure, scalable, and flexible AI solutions enables it to meet the diverse needs of its global clients. By combining AI with human expertise, SDL ensures that translations are not only accurate but also culturally relevant and contextually appropriate.

Lionbridge [198], another key player in the translation industry, utilizes AI to optimize translator assignments and improve content quality. The company has developed tools such as the Domain Detector and Customer Affinity, which use machine learning to match translators with projects based on their expertise and previous experience. Lionbridge's approach emphasizes a balance between automation and human oversight, ensuring that AI-driven processes enhance rather than diminish the role of skilled translators. This strategy enables Lionbridge to deliver high-quality translations efficiently while maintaining the integrity of the original content.

We also see Elite Translations Africa [199]using AI to preserve their African heritage, which extends beyond automating text translation to include considerations of benefiting cultural preservation and intellectual property protection. While companies like TransPerfect, SDL, and Lionbridge are committed to transparency in their AI processes, ensuring that content creators are compensated fairly and that translations adhere to ethical standards.

As AI continues to transform the translation industry, companies are exploring innovative applications to enhance accessibility and user engagement. For example, AI-powered tools are being used to streamline the production of multilingual content, improve real-time translation capabilities, and support complex localization projects. These advancements demonstrate how AI can be leveraged to meet the growing demand for high-quality, culturally sensitive translations in a globalized world.

### 3.6. Construction & Urban Planning

Collected policies from various construction & urban planning firms highlight the growing adoption of AI technologies to enhance productivity, efficiency, safety, and decision-making across the industry. Several companies, including Bechtel [200], Turner Construction [201], Hyundai E&C [202], and Shimizu Corporation [203], are leveraging AI and machine learning to optimize construction processes, such as project scheduling, safety management, and structural design.

Innovation is a key theme throughout the policies and guidelines, with companies like VINCI [204], Bouygues [205], and Grupo ACS [206] establishing dedicated platforms and initiatives to drive AI-driven innovations across their subsidiaries. These innovations span various applications, including predictive maintenance, virtual and augmented reality, and geolocation data collection. They also emphasize the importance of AI in addressing the construction industry's productivity challenges. Lendlease [207] highlights how AI can improve the quality of contract bids and streamline construction processes, ultimately leading to increased productivity and efficiency. Lastly, the UAE National Strategy for AI [208] provides an overarching framework for the development and adoption of AI across various sectors, including construction and urban planning. The strategy outlines the UAE's vision to become a global leader in AI by 2031, focusing on key enablers such as talent attraction, data infrastructure, and responsible governance.

In conclusion, the construction and urban planning industry is increasingly embracing AI technologies to tackle challenges related to productivity, safety, and efficiency. Companies are investing in innovation and collaborating with stakeholders to drive the responsible adoption of AI, while national strategies, such as the UAE's, provide a supportive framework for the industry's transformation.

### 3.7. Consulting & Management

The policies and guidelines from various consulting and management companies, including McKinsey [209], Deloitte [210], TCS [211], Infosys [212], KPMG [213], Clearsight Advisors [214], Falconi [215], Nous Group [216], and Africa International Advisors Group (AIA) [216], highlight the growing adoption and impact of AI and generative AI in business processes and decision-making. These companies are leveraging AI to enhance productivity, drive innovation, and deliver value to their clients.

A common theme across the collection is the importance of responsible AI adoption. Companies are developing guidelines and frameworks to ensure AI is implemented transparently, ethically, and in compliance with regulations [211], [212]. TCS, for example, offers a comprehensive portfolio of services covering the entire AI lifecycle, helping enterprises implement AI solutions in an unbiased and trustworthy manner. Another key aspect is the integration of people and technology. AIA's Futures Practice emphasizes the importance of cultivating an environment for people to adapt and excel in the future while harnessing the potential of emerging technologies [216]. Nous Group showcases their use of generative AI to support consulting work [217].

The collection also highlights the transformative potential of AI across various sectors. KPMG analyzes the impact of ChatGPT on businesses, emphasizing the need for top managers to address the potential of generative AI urgently to harness its benefits and avoid being left behind by competitors. Falconi presents a case study of a client who achieved significant improvements in production performance through the application of AI, demonstrating the promising prospects of AI in driving operational excellence. Security emerges as another critical aspect of AI adoption. Accenture discusses the dual role of AI in both automating hacking and bolstering security, emphasizing the need for businesses to approach AI security with a multifaceted strategy, implementing robust measures, conducting risk assessments, and collaborating with experts to stay ahead of emerging threats.

In conclusion, the consulting and management companies featured in this collection are at the forefront of guiding their clients through the complexities of the evolving AI landscape. By promoting responsible AI adoption, integrating people and technology, harnessing the transformative potential of AI across sectors, and prioritizing security, these companies are helping businesses navigate the challenges and opportunities presented by AI and generative AI.

### 3.8. Design & Fashion Technology

The design and fashion industry collection highlights the growing adoption and impact of AI across various aspects of the industry, from product design and content creation to supply chain management and customer experience.

Nike [218] employs AI to enhance customer experience through personalized recommendations, supply chain optimization, and IT operations. The company collaborates with partners like Cognizant to modernize its infrastructure and drive connected commerce. Ralph Lauren [219] is testing generative AI across various business functions, including copy editing, graphics, and computer programming, to improve productivity and drive better outcomes. The company is also exploring the use of NFTs and Web3 technologies to engage with customers. Shiseido [220] is using AI to drive gradual transformation across its brands, focusing on consumer intimacy and data-driven decision-making. The company has established internal initiatives like the Shiseido Digital Centre of Excellence and SHISEIDO+ to foster a culture of innovation and upskill employees.

LVMH [221] has partnered with Stanford University's Institute for Human-Centered AI to explore AI applications in customer experience, product design, marketing, manufacturing, and supply chain management. The collaboration aims to develop human-centered AI solutions that complement human creativity and expertise. Kering [222] is leveraging AI across its value chain, from trend prediction and demand planning to supply chain optimization and pricing. The company has prioritized AI projects and established a dedicated AI team to drive innovation and efficiency.

Havaianas [223] utilized Sitation's AI-powered content creation tool, RoughDraftPro, to generate accurate and consistent product descriptions for its Amazon listings, overcoming challenges related to incomplete product records and inconsistent branding. Falabella [224] hired Amelia, a conversational AI assistant, to provide 24/7 IT support for its 100,000 employees, optimizing the handling of support tickets and freeing up resources for more complex issues.

Cotton On [225] uses Dash Hudson's visual AI technology to create high-performing content, leverage data-backed insights for influencer collaborations, and craft strategic campaigns across marketing channels. The AI-driven approach has significantly improved the brand's engagement rates and content creation process.

In conclusion, the design and fashion industry is increasingly embracing AI to drive innovation, improve operational efficiency, and enhance customer experiences. While challenges such as job displacement and bias in AI models persist, companies are collaborating with academic institutions and technology partners to develop responsible and human-centered AI solutions that complement human creativity and expertise.

### 3.9. Entertainment & Game Development

The entertainment and game development collection highlights the growing adoption and impact of AI across various aspects of the industry, from game design and content creation to player experience and safety [226], [227], [228].

Electronic Arts [229] explores how AI is transforming entertainment and culture, focusing on the ethical and aesthetic considerations when using AI in social contexts. The company emphasizes the importance of purpose-driven, sustainable outcomes and responsible AI adoption. Tencent [230] applies AI across its businesses, including content recommendation, social interactions, and gameplay experience. The company also pursues key applications in industries such as medical, agriculture, industrial, and manufacturing, aiming to help enterprises achieve digital upgrades through AI. Square Enix's President [231] discusses the company's AI initiatives, highlighting the potential of generative AI to reshape content creation and fundamentally change programming processes. The company is investing in AI, blockchain entertainment, and the cloud to adapt to the changing business environment and drive innovation. Ubisoft's Ghostwriter [232], an in-house AI tool, assists scriptwriters by generating first drafts of barks, allowing them to focus on polishing the narrative. The tool demonstrates how AI can augment human creativity and streamline game development processes.

Supercell [233] is backing a venture capital fund dedicated to investing in early-stage AI startups. The fund, run by Air Street Capital, aims to identify and nurture AI companies that are developing innovative business models and solutions. Wildlife [234] prioritizes player safety by using AI scanning to monitor user-generated content and partnering with law enforcement and child protection agencies when necessary. The company's Trust and Safety team ensures the well-being of players through a combination of AI systems and human review. Globant's AI Manifesto [235] outlines the company's principles for responsible AI adoption, emphasizing augmented intelligence, respectful data, fairness, transparency, social contribution, and sustainable AI. The manifesto also lists applications that Globant will not support, such as misinformation, malicious use, and reckless AI.

In conclusion, the entertainment and game development industry is increasingly embracing AI to drive innovation, enhance player experiences, and streamline production processes. While challenges related to responsible AI adoption and ethical considerations persist, companies are collaborating with partners and developing guidelines to ensure the safe and beneficial integration of AI into their products and services.

### 3.10. Journalism & News Media

The journalism and news media industry is increasingly adopting AI to innovate, improve efficiency, and enhance audience engagement. AI is being used to generate articles, particularly for hyperlocal news [236], [237], personalize content recommendations, enhance accessibility, and improve user experience [237], [238]. However, human oversight remains crucial for maintaining accuracy and journalistic standards [237].

To ensure responsible AI use, many organizations have developed principles and guidelines focusing on fairness, security, privacy, intellectual property rights, human oversight, and accountability [239], [240], [241]. Data protection and regulatory compliance are significant challenges [241], [242], with governments and organizations working to implement safeguards and policies [242].

News organizations are collaborating with technology companies, academic institutions, and other stakeholders to advance AI capabilities and address challenges [238], [243]. While AI is seen as a powerful tool, there's an emphasis on maintaining human creativity, insight, and oversight in news production [237], [239], [241]. As AI evolves, news organizations must adapt to a complex regulatory landscape [241], [242], balancing AI benefits with core journalistic values of accuracy, transparency, and public trust.

### 3.11. Pharmaceutical Research & Development

The pharmaceutical industry is increasingly adopting AI to innovate across drug discovery, development, manufacturing, supply chain management, and patient care. AI is being used to accelerate drug discovery and development [244], [245], [246], [247], identify drug targets, generate molecular structures, predict drug-target interactions, and optimize clinical trials [246], [247]. Companies are establishing ethical principles and governance structures to ensure responsible AI use, focusing on fairness, transparency, accountability, privacy, security, and human oversight [244], [245], [248], [249], [250], [251].

The industry emphasizes using AI to augment human capabilities rather than replace them [244], [248], [249], [251], applying it to improve patient care, enhance diagnosis and treatment, and enable personalized medicine [244], [247], [252]. Companies are also addressing the environmental impact of AI [249], [251] and partnering with sustainable technology platforms.

As pharmaceutical companies navigate this complex landscape, they are developing governance frameworks, investing in employee training, and collaborating with stakeholders to ensure responsible and sustainable AI deployment in healthcare.

### 3.12. Social Media & Telecommunications

Social media and telecommunications companies are establishing ethical AI principles and guidelines. LinkedIn focuses on economic opportunity, trust, fairness, inclusion, transparency, and accountability [253]. WPP has developed comprehensive AI policies and ethics principles [254]. Privacy and data protection are prioritized, as seen in X's policy on synthetic media [255] and Skype's translation feature [256]. Transparency and explainability are key concerns, with companies like Spotify [257] and LinkedIn [253] striving to explain their AI systems clearly. Addressing AI biases and promoting fairness is a common theme [253], [255]. Companies are leveraging AI to enhance customer experiences, including Spotify's personalized recommendations [257] and Mercado Libre's e-commerce enhancements [258]. AI is also being used to boost operational efficiency and drive innovation. Examples include Alibaba Cloud's Tongyi Qianwen model [259], América Móvil and Carlos Slim Foundation's health monitoring tools [260], and Accel's focus on AI applications for business productivity [261].

In conclusion, the industry trend is towards comprehensive AI policies that address ethical concerns, promote responsible development, and leverage AI's potential while balancing innovation with necessary regulations.

### 3.13. Advertising & Marketing

AI is recognized as a transformative force in the advertising and marketing sector . Companies like Ogilvy [262], Publicis Groupe [263], and LegalZoom [264] are leveraging AI to revolutionize their industries. There's a consensus that AI will augment rather than replace human workers, with Ogilvy [262] and Publicis Groupe [263] emphasizing AI's role in enhancing human creativity and productivity.

AI is enabling unprecedented personalization and efficiency at scale. Omnicom [265] and Havas Media Group [266] use AI for targeted marketing campaigns and optimized media buying. Companies like Dentsu [267] and Publicis Groupe [263] are integrating AI across multiple business functions. The impact on workforce dynamics is stressed, with the ABC resource guide [268] noting potential job automation and new job creation. Legal and regulatory aspects of AI are still evolving, as evidenced by Hakuhodo DY Holdings' cautious approach [269] and LegalZoom's beta launch of AI-assisted services [264].

In conclusion, organizations are embracing AI for its transformative potential while recognizing the need for ethical considerations, responsible use, and workforce adaptability. Ongoing collaboration and adaptation will be crucial as AI technology continues to evolve rapidly.

### 3.12.Legal Technology & Services

AI within legal technology and services focuses on ethical considerations and emerging regulations. A consistent theme is balancing AI's transformative potential with responsible implementation. Thomson Reuters [270], Rocket Lawyer [271], and Demarest Advogados[272] emphasize key ethical principles for AI development and use, including beneficence, non-maleficence, autonomy, justice, and explicability. These inform practical guidelines for AI workplace policies and governance structures outlined by Cyril Amarchand & Mangaldas, Rocket Lawyer, and Allen & Ovary [271], [273], [274].

Zegal and Allen & Ovary highlight AI's impact on legal practices, revolutionizing areas like legal research, document review, and contract analysis [274], [275]. While noting efficiency gains, they stress the need for human oversight. Recommendations for responsible AI practices across multiple documents include developing AI-focused organizational cultures, conducting impact assessments, ensuring diverse teams, providing staff training, creating robust data policies, and implementing ongoing auditing processes [271], [272], [273]. Others discuss the emerging regulatory landscape for AI, including the EU AI Act and proactive efforts by companies to establish AI governance policies. This reflects a growing trend towards "responsible AI" as global regulatory scrutiny increases [276], [277], [278], [279]. Overall, the collection emphasizes AI's power to revolutionize industries, particularly law, while stressing the critical importance of ethical considerations and robust governance structures.

## IV. Quantitative Findings and Resultant Patterns

Our analysis employed text mining techniques, ranging from tokenization to visualization, to thoroughly explore AI guidelines across diverse industries. We conducted a Qualitative Semantic Analysis to identify ten key concepts within fourteen industrial sectors, focusing on both frequently mentioned terms (e.g., 'content,' 'data,' and 'risk') and those of unique relevance to specific sectors (e.g., conflict, fashion, treatment). Following this, a TF-IDF Heatmap Analysis measured the significance of terms within each industry, producing a cosine similarity matrix to visually compare thematic alignment across sectors. This integrated approach of qualitative depth and quantitative rigor illuminated shared priorities and sector-specific focuses, offering a foundational framework for enhancing and refining AI guidelines.

### 4.1. Qualitative Semantic Analysis

We conducted a qualitative semantic analysis by compiling key concepts from our literature review and identifying the ten most significant concepts for each of the fourteen categories. Each bar chart displays the frequency of these ten terms across the fourteen industry sectors. The selection of these key concepts was based on two criteria: their frequency of mention in the guidelines, indicating a consensus on their importance (e.g., 'content,' 'data,' and 'risk'), and their unique occurrence in the discourse, highlighting their specific significance despite their contextual nature (e.g., conflict, fashion, treatment). Figure 1 shows the frequency analysis of these qualitative findings.

**Figure 1: The Frequency of Key Concepts in the Major Nine Themes**

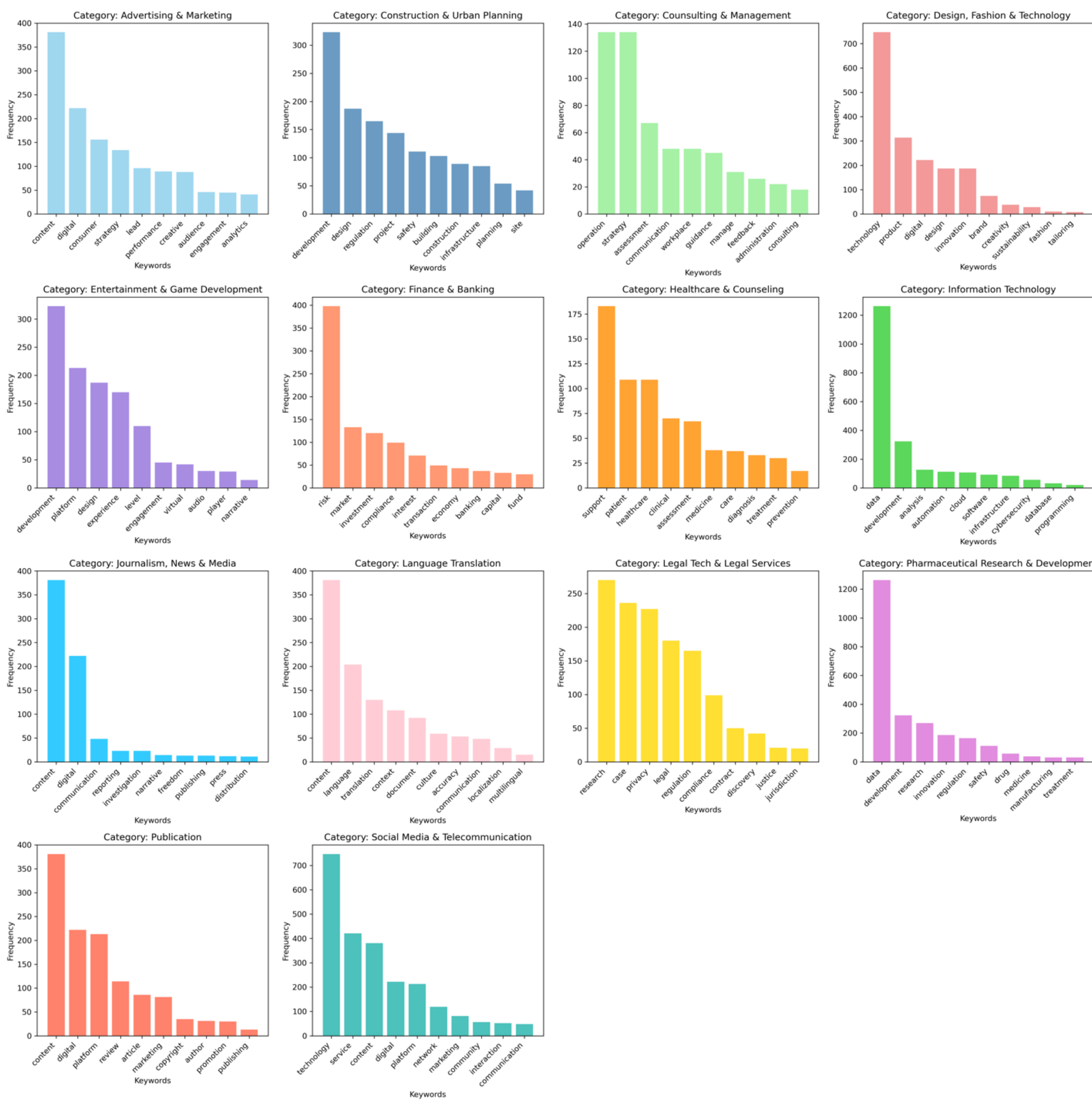

## 4.2. Industry-Wise Sankey Diagram Keyword Co-Occurrence Analysis

We conducted a quantitative semantic analysis using the top 10 key concepts from Figure 1 to match relevant cooccurring words and usage patterns from our guideline texts overlaps across 14 distinct industries using Sankey diagrams. These diagrams represent the flow and thematic alignment of AI-related guideline topics, highlighting the proportional emphasis placed by each industry areas of focus.

Each industry-specific Sankey diagram captures the distribution and flow of key word co-occurrences across various thematic categories, such as ethical complexities, innovation, decision-making paradigms, and collaborative creativity highlighted by our industry specific keywords. The diagrams also reveal commonalities and differences in priorities, providing a comparative perspective on how sectors integrate AI guidelines into their operational frameworks.

# Figures 2: Industry-Wise Sankey Diagram Analysis for AI Integration Guidelines

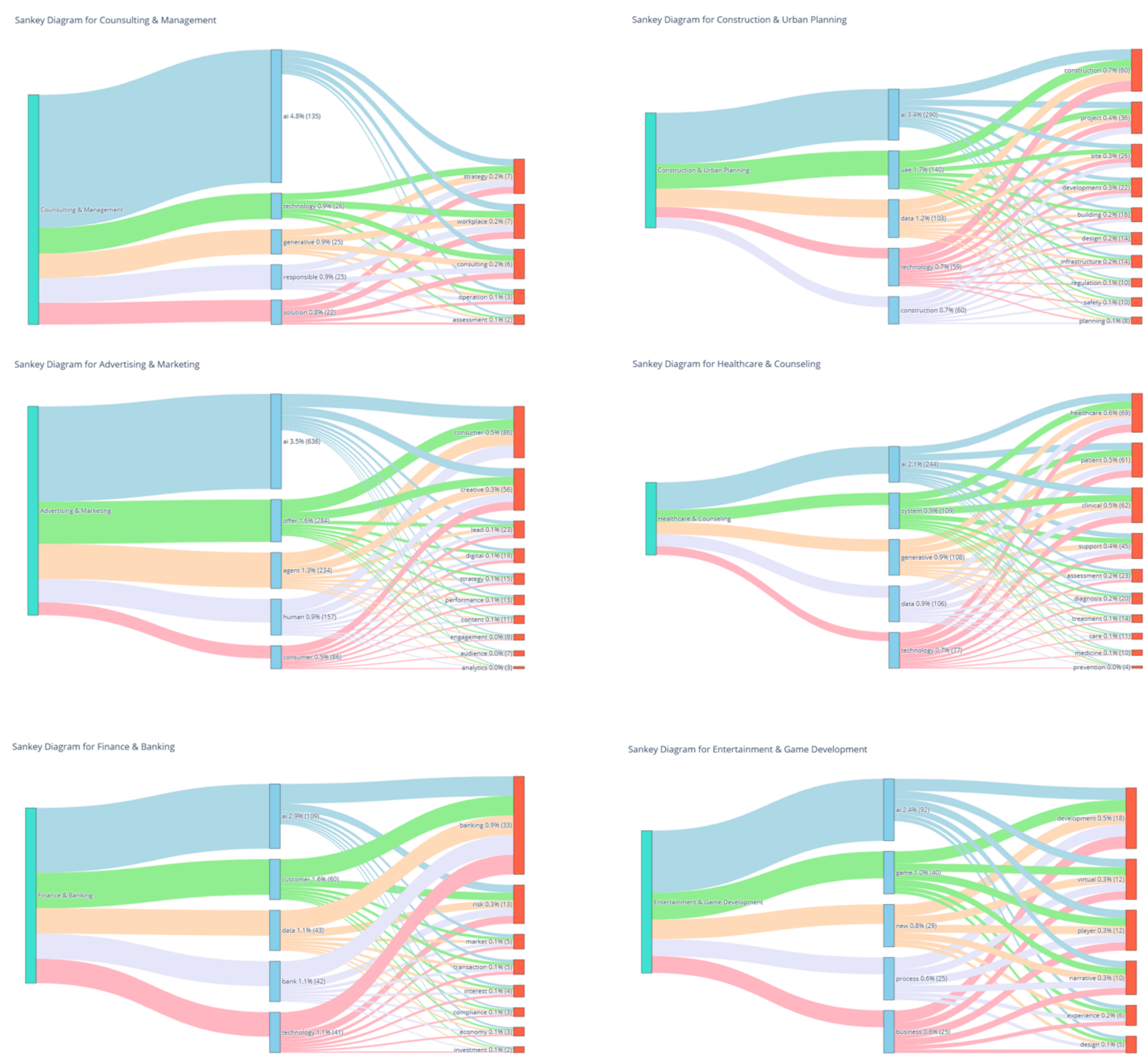

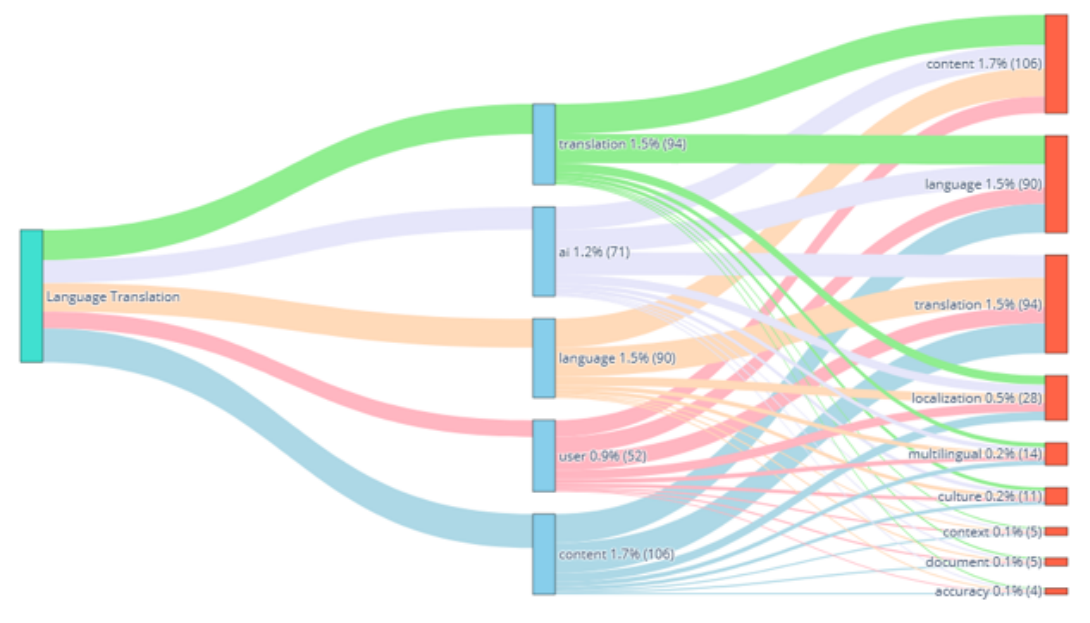

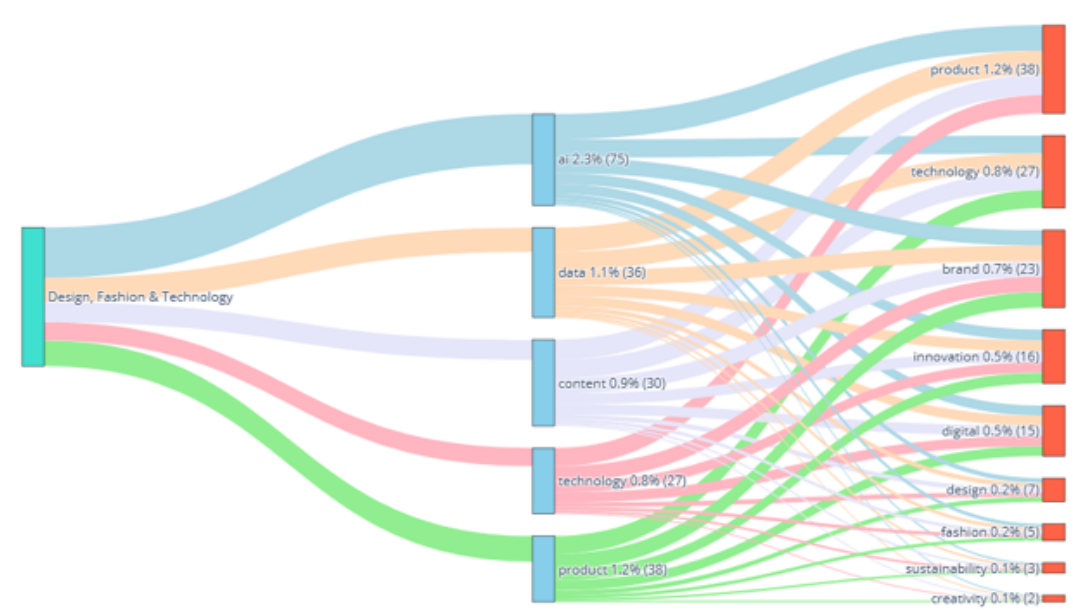
Sankey Diagram for Design, Fashion & Technology
Design, Fashion & Technology
ai 2.3% (75)
data 1.1% (36)
content 0.9% (30)
technology 0.8% (27)
product 1.2% (38)
product 1.2% (38)
technology 0.8% (27)
brand 0.7% (23)
innovation 0.5% (16)
digital 0.5% (15)
design 0.2% (7)
fashion 0.2% (5)
sustainability 0.1% (3)
creativity 0.1% (2)

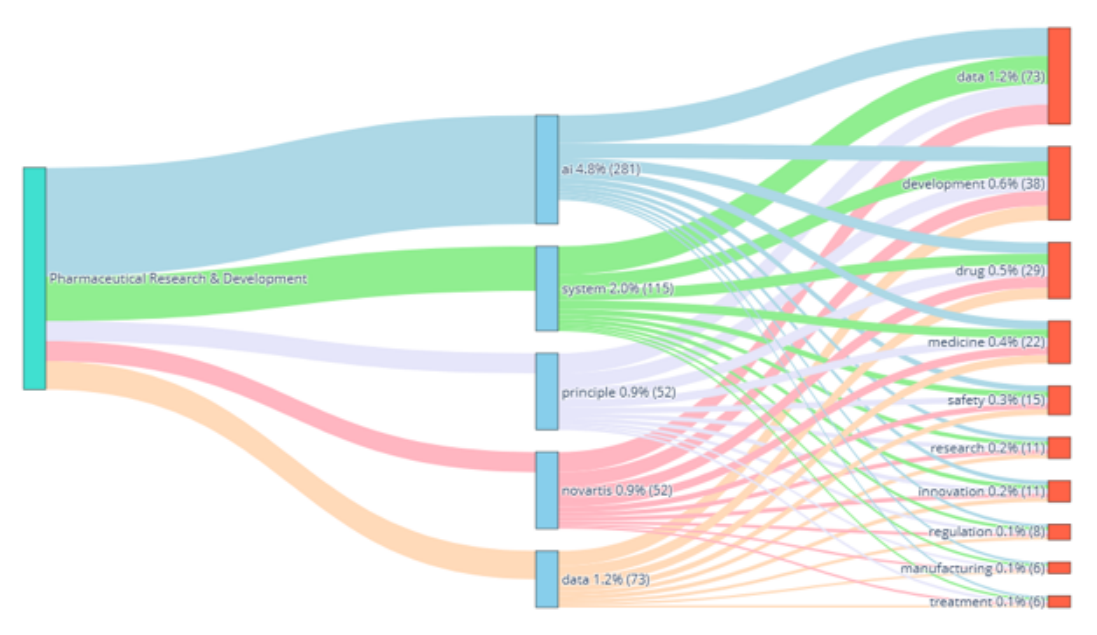

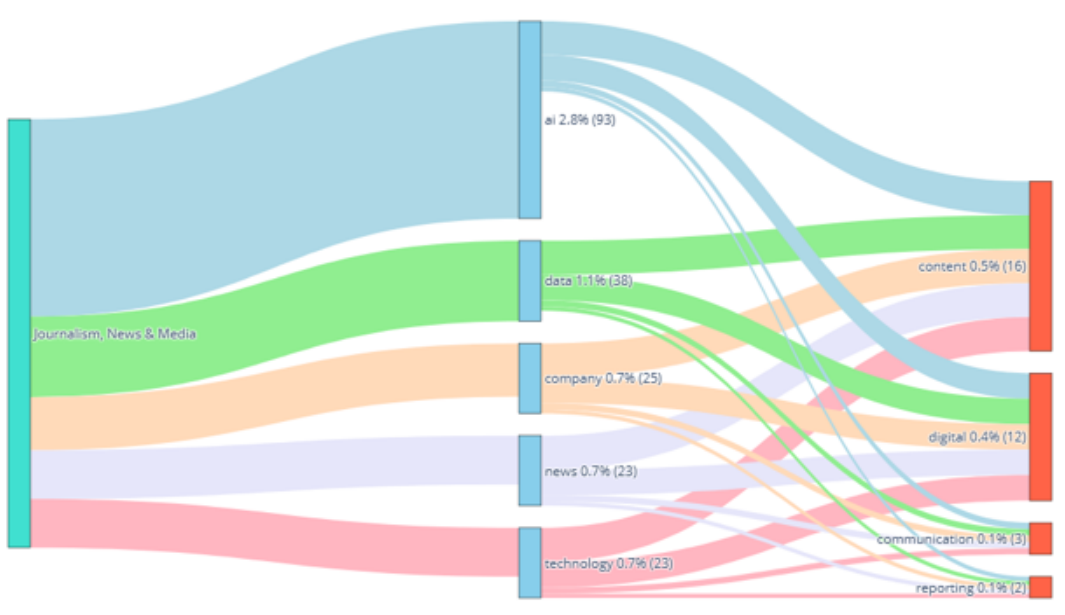
Sankey Diagram for Journalism, News & Media
Journalism, News & Media
ai 2.8% (93)
data 1.1% (38)
company 0.7% (25)
news 0.7% (23)
technology 0.7% (23)
content 0.5% (16)
digital 0.4% (12)
communication 0.1% (3)
reporting 0.1% (2)

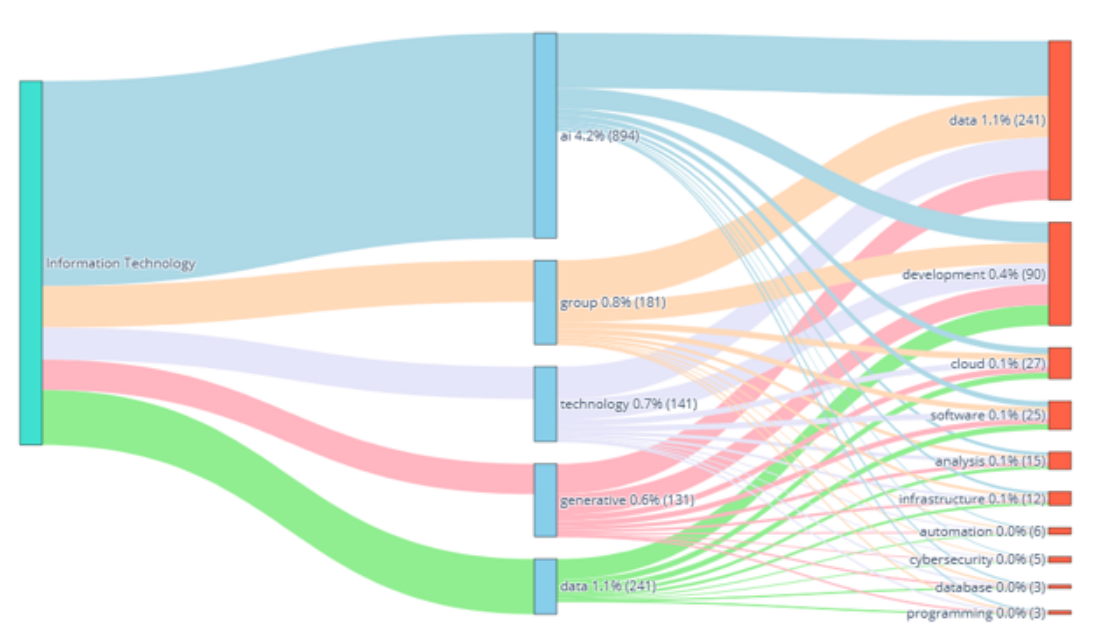
Sankey Diagram for Information Technology
Information Technology
ai 4.2% (894)
group 0.8% (181)
technology 0.7% (141)
generative 0.6% (131)
data 1.1% (241)
data 1.1% (241)
development 0.4% (90)
cloud 0.1% (27)
software 0.1% (25)
analysis 0.1% (15)
infrastructure 0.1% (12)
automation 0.0% (6)
cybersecurity 0.0% (5)
database 0.0% (3)
programming 0.0% (3)

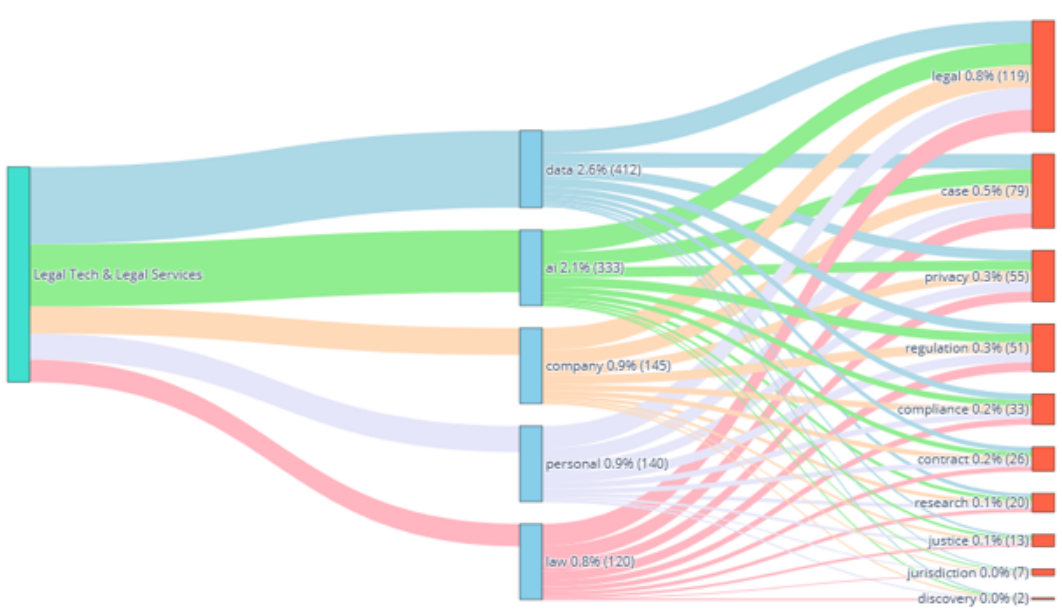
Sankey Diagram for Legal Tech & Legal Services
Legal Tech & Legal Services
data 2.6% (412)
ai 2.1% (333)
company 0.9% (145)
personal 0.9% (140)
law 0.8% (120)
legal 0.8% (119)
case 0.5% (79)
privacy 0.3% (55)
regulation 0.3% (51)
compliance 0.2% (33)
contract 0.2% (26)
research 0.1% (20)
justice 0.1% (13)
jurisdiction 0.0% (7)
discovery 0.0% (2)

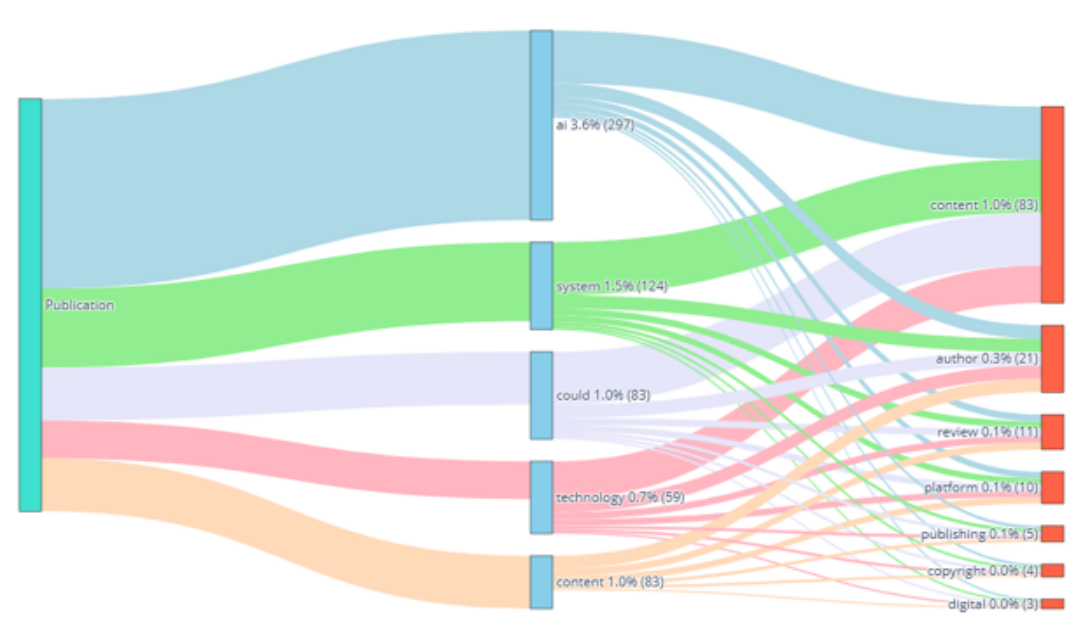
Sankey Diagram for Publication
Publication
ai 3.6% (297)
system 1.5% (124)
could 1.0% (83)
technology 0.7% (59)
content 1.0% (83)
content 1.0% (83)
author 0.3% (21)
review 0.1% (11)
platform 0.1% (10)
publishing 0.1% (5)
copyright 0.0% (4)
digital 0.0% (3)

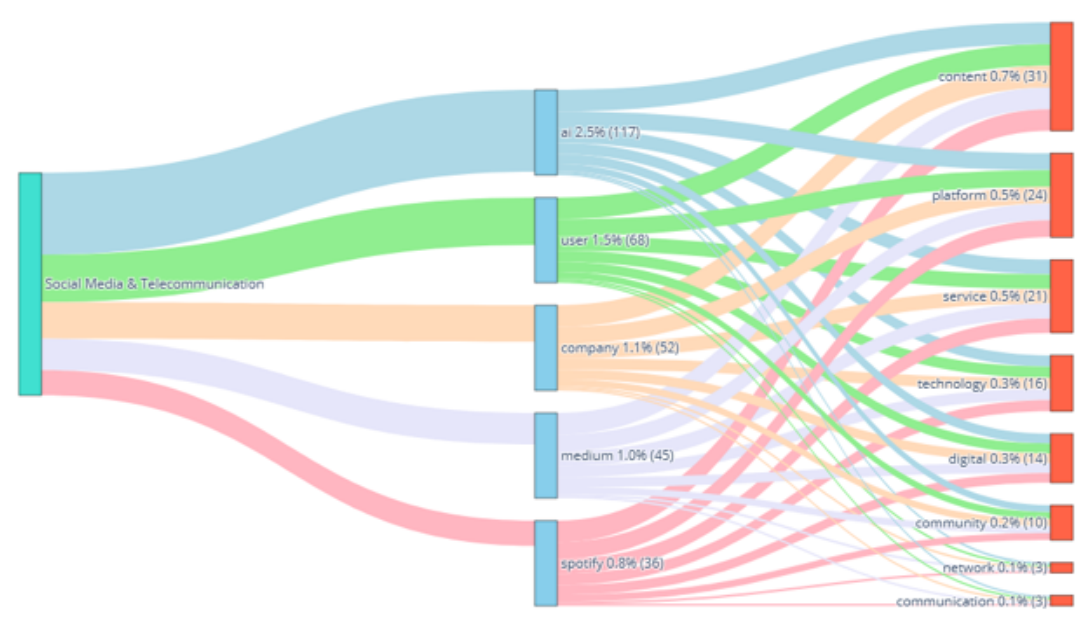
Sankey Diagram for Social Media & Telecommunication
Social Media & Telecommunication
ai 2.5% (117)
user 1.5% (68)
company 1.1% (52)
medium 1.0% (45)
spotify 0.8% (36)
content 0.7% (31)
platform 0.5% (24)
service 0.5% (21)
technology 0.3% (16)
digital 0.3% (14)
community 0.2% (10)
network 0.1% (3)
communication 0.1% (3)

**Figures 2** illustrate these Sankey diagrams for industries ranging from healthcare and finance to journalism and legal services. By mapping these flows, we gain insights into the unique thematic concentrations and shared concerns in AI adoption across sectors. This analysis underscores key areas for future development, including addressing ethical dilemmas, ensuring fairness, and fostering innovation while balancing traditional approaches with emerging technologies.

## V. Discussion and Synthesis

In this section, we also synthesize these findings to propose recommendations for refining industry-specific AI integration guidelines. This includes fostering cross-sector collaboration to address shared challenges, enhancing safety and responsibility protocols, and navigating the complex interplay between technological optimism and ethical caution in AI adoption.

### 5.1. Synthesis of Quantitative and Qualitative Insights

Our semantic analysis of key concepts, combined with the keyword frequency and The Sankey diagrams provided visualize keyword co-occurrence percentages with other terms across 14 industries, reveals that certain terms, despite appearing less frequently, are essential for future guidelines to offering insights into thematic overlaps and industry-specific focal points of GAI and LLMs.

Privacy is a high-priority concept across multiple industries, with Legal Tech & Legal Services showing 227 instances of "privacy," underlining its critical role in protecting user information. The visualization confirms this, showing "privacy" as one of the more frequently co-occurring words with "data" (2.6%, 412 occurrences) and "AI" (2.1%, 333 occurrences) in the legal domain, which highlights its prominence in the industry's focus on compliance and protection. However, disclosure, with only 14 mentions across all documents, is underrepresented, despite its significance in promoting transparency. The Sankey diagram for Journalism, News & Media also reflects limited emphasis on transparency-related terms, with "communication" and "reporting" appearing weakly (0.1%, 3 occurrences). Emphasizing "disclosure" in all industry guidelines would help users understand AI capabilities and limitations, fostering accountability and informed decision-making.

In the Ethical Complexities & Human-Centric Usage category, “ethical” appears frequently, with 216 instances across industries like Counseling & Management and Healthcare & Counseling, emphasizing its importance. The diagrams show "ethical" connecting strongly to “AI” (4.8%, 135 occurrences) in Counseling & Management, reflecting its centrality in AI-driven solutions. In contrast, “human-centric” is mentioned only twice, suggesting a gap in guidelines promoting inclusivity and accessibility. The Social Media & Telecommunication sector, with its strong user interaction (381 mentions of "content" and 52 mentions of "interaction"), could benefit from a greater emphasis on "human-centric" approaches. Expanding guidelines to promote inclusivity on high-interaction platforms would enhance trust and accessibility, particularly as the diagrams reveal a focus on "user" (1.5%, 68 occurrences) and "service" (0.5%, 21 occurrences) but limited emphasis on equity-driven practices.

In the Balancing Innovation & Integrity category, “integrity” appears frequently (211 instances), especially in sectors where ethical standards are essential, like Pharmaceutical Research & Development. The visualization reinforces this, showing "integrity" linked strongly to "AI" (4.8%, 281 occurrences) and "system" (2.0%, 115 occurrences), reflecting its critical role in ensuring ethical drug development. Meanwhile, “alternative methods” are referenced only 47 times across industries, indicating an area where innovation could be encouraged. Sectors like Education and Healthcare Counseling could benefit from exploring "alternative methods" such as adaptive learning and AI-driven diagnostic tools, balancing innovation with ethical standards. Additionally, the analysis reveals a low frequency of critical concepts like misinformation (19 instances) and skepticism (only 5 mentions), particularly in Journalism, News & Media and Social Media & Telecommunication. This is reflected in the weak connections to terms like "compliance" and "truthfulness" in their respective Sankey diagrams. Expanding guidelines on misinformation detection and skepticism could strengthen truthfulness and risk mitigation strategies, fostering trust in AI systems. This is especially relevant in Journalism, News & Media, where "content" appears prominently (2.8%, 93 occurrences), but guidelines around responsible information sharing remain limited.

In the Collaborative Creativity & Co-Designing category, assistance is frequently noted (41 instances), whereas democratization appears just once across all guidelines, highlighting a significant gap. The diagrams emphasize this gap, particularly in Entertainment & Game Development, where "platform" is referenced 213 times, showing the industry's reliance on AI-driven systems. However, democratization efforts such as open-source tools and collaborative development platforms are not evident. Broadening these efforts could make AI more accessible and foster co-design with diverse user groups. Initiatives like educational resources and co-creation practices would enhance accessibility and equity in AI innovation.

Predictive analytics, referenced in Finance & Banking (120 mentions of "investment" and 133 of "market"), plays a significant role in empowering decision-making. The Sankey diagram for Finance & Banking confirms this by showing strong links between "AI" (2.9%, 109 occurrences) and investment-related terms like "banking" and "market." Leveraging "predictive analytics" in educational settings or data-driven industries could provide insights for better resource allocation and student support, optimizing outcomes. For instance, Healthcare Counseling, with 183 instances of "support" and 109 of "patient," could integrate predictive tools for patient care and resource management. Additionally, promoting critical thinking in AI-powered learning tools encourages analytical reasoning and problem-solving skills, benefitting students and professionals alike across fields.

Lastly, the analysis shows that support-focused guidelines are more prevalent, while those for employees and management are comparatively underrepresented. For example, Counseling & Management guidelines contain 134 instances of "strategy," showing a strong emphasis on organizational direction. However, the diagrams reveal limited co-occurrences of terms like "leadership" (0.1%, 2 occurrences), indicating a gap in addressing educational and managerial roles. Developing discipline-specific guidelines tailored to fields like Healthcare (183 instances of "support" and 109 of "patient") and Legal Services (227 mentions of "privacy") could ensure AI usage aligns with each field’s unique challenges and ethical considerations. Working with domain experts to refine industry-specific practices, as visualized in the Sankey diagrams, will enable responsible and effective AI implementation across diverse sectors.

### 5.2. Dynamic and Modular Co-Creation: A Framework for Adaptive AI Governance

The complexity of AI technologies necessitates a shift from static, top-down guidelines to a dynamic and modular co-creation framework. This approach emphasizes continuous collaboration among developers, regulators, users, and ethicists to treat guidelines as "living documents" that evolve with real-world feedback and technological advancements. AI-driven tools, such as sentiment analysis, can further support updates by identifying emerging gaps and concerns, fostering trust and accountability.[280][281]

A key component is AI-enhanced meta-governance, where AI audits evaluate adherence to ethical principles by analyzing public outputs and exposing inconsistencies. This incentivizes transparent and actionable guidelines. Complementing this, modular governance structures combine universal ethical principles with sector-specific and operational standards, allowing industries to adapt efficiently while maintaining accountability. By enabling organizations to "opt-in" to tailored modules, this flexible system ensures guidelines remain relevant and practical across diverse contexts.

Together, continuous co-creation, meta-governance, and modular structures form a responsive framework that addresses the shortcomings of static governance models. This integrated approach promotes transparency, collaboration, and accountability, paving the way for ethical and adaptive AI deployment.

### 5.3. Embedding Values and Incentivizing Human-Centric AI Development

AI guidelines often emphasize human-centric design superficially, overlooking the critical role of embedding values during the pre-design and in-design phases. To address this, we propose mandating participatory design practices involving diverse stakeholders—end-users, ethicists, and domain experts—early in development to ensure AI systems align with societal and cultural contexts. For example, healthcare AI should integrate cultural and emotional intelligence metrics to deliver not just accurate but empathetic and context-sensitive recommendations.

To further drive this commitment, governments and regulators should incentivize responsible innovation through grants, tax breaks, and certifications like "Ethical AI Leader" awards. These measures reward companies prioritizing bias mitigation, ethical innovation, and open-source contributions, fostering accountability and competition in ethical practices. By embedding values early and offering strong incentives, this integrated approach ensures AI systems enhance societal well-being while promoting a culture of ethical responsibility and long-term innovation.[282]

### 5.4 AI-Driven Mediation and Sandbox Models for Inclusive and Evidence-Based Governance

Addressing diverse stakeholder priorities and the societal impacts of AI requires innovative governance strategies that combine AI-driven mediation with sandbox experimentation. Generative AI tools can act as "negotiation assistants," helping developers, regulators, and civil society reconcile competing goals by simulating scenarios, quantifying trade-offs, and proposing solutions such as anonymization protocols or tiered data access. These tools foster balanced decision-making by transforming abstract debates into actionable insights while highlighting AI's potential to facilitate consensus.

Complementing this, sandbox environments provide controlled spaces to test AI systems and guidelines under realistic conditions. Stakeholders can evaluate tools across diverse user populations and regulatory contexts, refining both technology and governance mechanisms before large-scale deployment. For instance, a healthcare sandbox could assess AI compliance and effectiveness across varying demographics and resource levels. Sandboxes also support testing novel oversight models, ensuring governance frameworks are evidence-based and adaptable. By integrating mediation tools and sandbox models, this approach bridges conflicting interests and proactively evaluates guidelines, creating a governance framework that is inclusive, accountable, and aligned with real-world demands.

### 5.5. Dynamic Regulation in Healthcare and Pharmaceuticals: Balancing Innovation and Safety

The healthcare and pharmaceutical industries are often constrained by rigid regulatory frameworks, which can stifle the rapid adoption of AI. While stringent regulations prioritize patient safety and compliance, they also inhibit innovation and the timely integration of cutting-edge technologies. One approach could be to implement adaptive regulatory frameworks that dynamically adjust compliance requirements based on real-time risk assessments of AI systems. For example, AI-driven tools for low-risk applications like patient scheduling could face relaxed oversight compared to high-risk systems like AI-assisted diagnostics. However, the trade-off here is ensuring that adaptive frameworks do not inadvertently compromise safety or create regulatory loopholes. The key lies in developing AI systems that not only meet adaptive regulatory requirements but also enhance regulators' capacity to monitor evolving risks using AI-driven predictive models.

### 5.6. Augmentation-First Guidelines for Creative and Linguistic AI

In creative and linguistic fields, AI is best positioned as a tool for augmentation rather than replacement. Tasks like translating literature, where cultural nuance and contextual relevance are crucial, highlight the insufficiency of fully automated AI outputs. Industries should adopt Augmentation-First Guidelines that mandate human review and refinement of AI-generated content. For instance, AI-translated works should be edited by linguistic experts to preserve cultural integrity and idiomatic expressions. Moreover, continual training pipelines are necessary for AI models to keep pace with evolving language trends, including slang and colloquialisms. While this process might slow deployment timelines, it prioritizes quality and sensitivity, ensuring AI enhances human creativity and understanding rather than producing substandard or culturally insensitive results.

### 5.7. Cautious Innovation in Banking and Beyond

The banking sector's measured approach to AI adoption offers valuable lessons for other high-stakes industries. Banks often pilot new technologies rigorously, performing risk assessments and consulting stakeholders before full-scale deployment. This Cautious Innovation Model should be formalized and extended to fields like healthcare, where patient outcomes are directly affected. For example, healthcare providers could test AI diagnostic tools in controlled environments, gathering evidence of safety and efficacy before integrating them into routine care. Although this incremental process may slow innovation, it reduces unintended consequences and fosters stakeholder trust. Additionally, sharing findings from these pilots through open-access platforms would enable industries to collectively refine AI systems and guidelines, accelerating responsible and evidence-based innovation.

### 5.8. Transparent AI in Social Media: Trade-Off Between Privacy and Accountability

In the social media sector, transparency initiatives such as Explainable AI (XAI) aim to make algorithmic decision-making comprehensible to users. While transparency fosters trust and accountability, it may inadvertently conflict with privacy regulations such as GDPR and CCPA, as revealing too much information about algorithms might expose sensitive data or proprietary technologies. A solution could involve tiered transparency mechanisms, where different levels of algorithmic explanations are provided based on stakeholder roles. For example, general users could receive high-level summaries, while regulators and auditors access detailed algorithmic insights. However, this approach requires careful calibration to avoid undermining user privacy or exposing companies to competitive risks. Balancing these trade-offs would necessitate the creation of robust AI audit systems that ensure algorithmic accountability without compromising proprietary or personal data.

### 5.9. AI Innovation Hubs in Construction and Urban Planning: Balancing Sustainability and Costs

AI adoption in construction and urban planning has focused heavily on efficiency and safety, yet the environmental impact of AI technologies is often overlooked. For instance, the energy consumption of AI systems in predictive maintenance or structural design can be substantial. To address this, industries could establish AI Innovation Hubs that prioritize sustainable AI development, including energy-efficient algorithms and carbon-neutral data centers. While such hubs could drive environmentally friendly practices, they may also introduce cost burdens for smaller firms with limited resources. A potential solution is creating public-private partnerships that subsidize the initial costs of adopting sustainable AI practices, ensuring broader participation without sacrificing innovation or sustainability goals.

### 5.10. Human-AI Collaboration in Design and Entertainment: Bridging Creativity and Ethics

In creative sectors like design, fashion, and entertainment, AI tools are often celebrated for enhancing productivity and innovation but criticized for potentially eroding human creativity. To bridge this gap, a novel proposal is to establish Human-AI Collaboration Guidelines that define ethical boundaries for AI use in creative processes. For instance, guidelines could mandate that AI-generated designs are clearly labeled and that human creators retain intellectual property rights over AI-augmented work. While these measures can protect human creativity and ethical integrity, they may also limit the full potential of AI's capabilities. Striking a balance requires fostering an ecosystem where AI serves as a tool for augmenting human creativity rather than replacing it, supported by training programs that equip creators to work effectively with AI.

### 5.11. Ethical Auditing in Journalism and Language Translation: Democratizing Oversight

Journalism and language translation industries are uniquely positioned to democratize the oversight of AI systems due to their reliance on public trust and cultural sensitivity. An helpful approach could be to develop community-driven ethical auditing platforms, where linguists, journalists, and the public can actively contribute to identifying ethical lapses in AI systems. For example, users could flag instances of biased translations or misinformation in AI-generated news articles. While this democratized approach fosters inclusivity and accountability, it may also lead to conflicts over what constitutes ethical behavior, particularly in culturally sensitive contexts. To mitigate these challenges, industries could establish consensus-driven review panels that validate flagged issues and incorporate them into AI system updates, ensuring that community feedback translates into actionable improvements.

### 5.12. Legal Technology and AI: Proactive Governance for Emerging Risks

In legal technology, the use of AI for tasks like document review and contract analysis offers tremendous efficiency gains but raises ethical concerns about bias and autonomy. A proactive governance approach would involve creating pre-emptive risk models that simulate the potential biases and ethical dilemmas of AI systems before they are deployed. For example, legal tech firms could use generative AI to simulate real-world use cases and identify scenarios where biases might emerge, such as favoring certain demographics in contract analysis. While this proactive approach can reduce downstream risks, it also requires significant investment in simulation tools and expertise. Collaboration with academic institutions and ethics researchers could help offset these costs while ensuring robust and impartial governance.

### 5.13. Adaptive Learning in Advertising and Marketing: Personalization Without Exploitation

AI-driven personalization in advertising and marketing often walks a fine line between delivering value to consumers and exploiting their data. Some adaptive learning models are proposed that allows consumers to set their own boundaries for how their data is used. However, implementing such models requires significant investment in user-friendly interfaces and robust data management systems. The trade-off lies in balancing user autonomy with the industry's need for data-driven insights, which could be addressed through transparent communication and incentivizing users to share data responsibly.

### 5.14. The Risks of Marketing Hype in AI Guidelines and Policies

AI guidelines and policy statements often exaggerate the capabilities and impacts of AI systems, driven by marketing agendas rather than evidence-based assessments. Companies frequently describe their AI tools as "revolutionary" or "transformative," promising universal solutions while failing to provide empirical support. For example, healthcare AI is often claimed to "transform global health outcomes," yet adoption and efficacy vary significantly across diverse and under-resourced settings. Similarly, financial institutions tout the precision of AI in fraud detection without transparent metrics to substantiate these claims.

This hype-driven approach creates unrealistic expectations, misallocates resources toward promotional efforts, and risks eroding public trust when exaggerated benefits fail to materialize.

Furthermore, a troubling disconnect exists between companies' claims of "responsible AI" and the actual implementation of ethical practices, with many systems lacking transparency or exhibiting bias.

To counter these issues, companies must adopt evidence-based reporting, distinguishing between potential and demonstrated AI capabilities. Transparent evaluations, independent audits, and third-party validations should become standard, ensuring that policy statements align with verifiable outcomes. By reducing hyperbolic rhetoric, organizations can build trust, credibility, and genuine progress in ethical AI deployment.

## VI. Concluding Remarks and Future Directions

The rapid evolution and integration of Generative AI (GAI) and Large Language Models (LLMs) have ushered in transformative opportunities across diverse industrial sectors while simultaneously introducing significant ethical, operational, and regulatory challenges. This study provided a comprehensive analysis of 160 guidelines and policy statements across 14 industrial sectors, offering critical insights into the governance of GAI and LLMs. Our findings reveal key themes such as the centrality of privacy, the underrepresentation of concepts like disclosure and human-centricity, and the sector-specific emphasis on integrity, innovation, and ethical accountability. These findings highlight the urgent need for robust, inclusive, and adaptive frameworks that address both the unique challenges and shared opportunities posed by these technologies.

The synthesis of qualitative and quantitative insights underscores the importance of moving beyond static governance models to dynamic, modular, and co-creative approaches that foster collaboration across stakeholders. By embedding human values into AI systems during the pre-design and in-design phases, industries can ensure that these technologies align with societal needs and ethical principles. Moreover, sector-specific guidelines, such as those emphasizing misinformation detection in journalism or augmentation-first practices in creative fields, demonstrate the necessity of tailoring governance strategies to the unique demands of each domain.

Despite the progress achieved through this analysis, gaps remain in the current understanding and governance of GAI and LLMs. Concepts such as democratization, alternative methods, and skepticism are underrepresented in existing guidelines, highlighting the need for future efforts to prioritize inclusivity, critical thinking, and innovation. Furthermore, the hype-driven narrative surrounding AI capabilities continues to obscure the practical limitations of these systems, necessitating a shift toward evidence-based evaluations and transparent reporting.

Looking ahead, future research and policy efforts must prioritize the development of adaptive and modular governance frameworks that evolve alongside technological advancements and real-world feedback. These frameworks should balance universal ethical principles with sector-specific modules, ensuring flexibility, accountability, and relevance across industries. Participatory design practices that actively involve diverse stakeholders, including end-users, ethicists, and domain experts, are crucial for embedding societal and cultural values into AI systems from the outset. Additionally, AI-driven mediation tools and sandbox testing environments can facilitate the reconciliation of competing stakeholder priorities and enable evidence-based, practical guidelines aligned with real-world demands.

Transparency and education must remain central to fostering trust and accountability. Explainable AI (XAI) initiatives and tiered transparency mechanisms can cater to diverse stakeholder needs, balancing privacy concerns with algorithmic accountability. To ensure sustainability, the establishment of AI Innovation Hubs that emphasize energy-efficient and environmentally friendly practices, supported by public-private partnerships, is imperative. Furthermore, tackling misinformation and curbing the exaggeration of AI capabilities require evidence-based evaluations and independent audits to align policy statements with verifiable outcomes. Together, these efforts will foster a governance ecosystem that promotes ethical, inclusive, and sustainable innovation, ensuring the responsible deployment of GAI and LLMs across diverse industries.

By addressing these future directions, researchers, policymakers, and industry leaders can collaboratively shape the next generation of GAI and LLM governance. Such efforts will ensure that these transformative technologies enhance human well-being, foster equitable access, and uphold the highest standards of ethical accountability. Ultimately, the responsible integration of GAI and LLMs into industry will depend not only on technological advancements but also on our collective commitment to fostering innovation that is both inclusive and sustainable.

### Acknowledgements

This research was funded by the National Science Foundation under grant number 2125858 and UT-Good Systems. The authors would like to express their gratitude for the foundation's support, which made this study possible. Furthermore, in accordance with MLA, we would like to thank OpenAI for its assistance in editing and brainstorming.